\pdfoutput=1
\documentclass[sigplan,screen,nonacm,twocolumn]{acmart}
\RequirePackage{etoolbox}

\providebool{fullversion}
\providebool{arxivversion}

\usepackage[utf8]{inputenc}
\usepackage[T1]{fontenc}
\usepackage[english]{babel}

\providecommand{\llbracket}{[\![}
\providecommand{\rrbracket}{]\!]}
\usepackage{tikz}
\usetikzlibrary{cd}
\usepackage{amsmath,amsthm}
\usepackage{mathpartir}
\usepackage{multicol}
\usepackage{mathtools}
\usepackage{tabularx}
\usepackage{booktabs}

\usepackage{hyperref}
\usepackage{cleveref}

\usepackage{cat}
\usepackage{grammar}
\usepackage{pl}
\usepackage{local-abbrv}
\usepackage{locallabel}
\usepackage{pftools}
\usepackage{macro}
\usepackage{xspace}

\newcommand{\grean}{Garlene\xspace} \newcommand{\garlene}{\grean}

\usepackage{listings}

\usepackage{color}
\definecolor{keywordcolor}{rgb}{0.7, 0.1, 0.1}   \definecolor{tacticcolor}{rgb}{0.0, 0.1, 0.6}    \definecolor{commentcolor}{rgb}{0.4, 0.4, 0.4}   \definecolor{symbolcolor}{rgb}{0.0, 0.1, 0.6}    \definecolor{sortcolor}{rgb}{0.1, 0.5, 0.1}      \definecolor{attributecolor}{rgb}{0.7, 0.1, 0.1} 

\definecolor{statehl}{gray}{0.90}
\newcommand{\hlline}[1]{{\setlength{\fboxsep}{1.5pt}\colorbox{statehl}{\ttfamily\small #1}}}

\ifbool{fullversion}{
\newcommand{\appref}[1]{\cref{#1}}
  \newcommand{\Appref}[1]{\Cref{#1}}
}{
\newcommand{\appref}[1]{the Appendix}
  \newcommand{\Appref}[1]{The Appendix}
}

\ifbool{arxivversion}{
\usepackage{caption}
}{
\usepackage[belowskip=-10pt,aboveskip=4pt]{caption}
\setlength{\intextsep}{15pt} }

\makeatletter
\ifbool{arxivversion}{}{
\let \MathparLineskip \mpr@lesslineskip }
\makeatother

\ifbool{arxivversion}{
  \newcommand{\vsquish}[1]{}
}{
  \newcommand{\vsquish}[1]{\vspace{-#1}}
}

\usepackage{tcolorbox}
\tcbuselibrary{skins,breakable}
  \newtcolorbox{result}{
    blanker,
    extras={interior engine=spartan},
    grow to left by=2pt,left*=0mm,
    grow to right by=2pt,right*=0mm,
    top=1mm,bottom=1mm,
    beforeafter skip balanced=0.1\baselineskip plus 2pt,
    breakable,
    colback=cyan!40
  }

\ifbool{fullversion}{
}{}

\title{\grean: Guarded Recursion in Lean}

\author[S. Stepanenko]{Sergei Stepanenko}
\authornote{This work was carried out while the author was affiliated with the IT University of Copenhagen.}
\orcid{0000-0002-7322-5644}
\affiliation{\institution{Aarhus University}
  \city{Aarhus}
  \country{Denmark}
}
\email{sergei.stepanenko@cs.au.dk}

\author[P. Bahr]{Patrick Bahr}
\orcid{0000-0003-1600-8261}
\affiliation{\institution{IT University of Copenhagen}
  \city{Copenhagen}
  \country{Denmark}
}
\email{paba@itu.dk}

\author[R.E. M{\o}gelberg]{Rasmus Ejlers M{\o}gelberg}
\orcid{0000-0003-0386-4376}
\affiliation{\institution{IT University of Copenhagen}
  \city{Copenhagen}
  \country{Denmark}
}
\email{mogel@itu.dk}

\begin{document}
\begin{abstract}

Extending the recursion principles of a formal system is an enticing
but dangerous endeavour with a well-documented history of leading to
consistency bugs. Nakano's guarded recursion is an elegant,
type-based approach to soundly extend type theory with a powerful
recursion principle. This makes guarded recursion useful for many
applications, from programming with infinite structures such as streams
to reasoning about advanced programming language features using
synthetic guarded domain theory. Sadly, guarded recursion is not
directly supported by any major interactive theorem prover, which
leaves users of guarded recursion with unmechanised pen-and-paper
proofs or mechanisations that depend on unmaintained theorem provers.

In this paper, we present an implementation of guarded recursion as an
embedded language in Lean consisting of a simply-typed lambda calculus
for definitions and a higher-order logic for reasoning. Using Lean's
excellent support for metaprogramming, our language allows users to
write guarded recursive definitions in an intuitive syntax and to
prove properties about them using a dedicated proof mode. We give our
language a presheaf model, which we use to prove the soundness of our
language and to allow users to export guarded recursive definitions
and their theorems into standard Lean developments. To demonstrate the
usefulness of our language, we present several case studies for
programming and reasoning with guarded recursion.
\end{abstract}

\maketitle
\section{Introduction}
\label{sec:intro}

Guarded recursion~\cite{Nakano:Modality} is a powerful framework for programming and reasoning with recursion in type theory. 
The central gadgets of guarded recursion are 1) the `later' modality mapping a type $A$ to the type $\tlater{A}$ of values of type $A$
delayed by one time step, 2) a fixed point operator $\fix : (\tlater A \to A ) \to A$, and 3) guarded recursive types, i.e., 
solutions to equations such as 
\begin{equation} \label{eq:untyped:lambda:model}
  D \iso \mathbb{N} + 1 + \tlater{D} + \tlater{(D \to D)}
\end{equation}
where the recursion variable $D$ occurs only under the modality $\tlater{}$, but is allowed to occur also in negative positions. 
One way to think of guarded recursion is as an abstract approach to step-indexing~\cite{AppelM01}, 
and this view is made precise in the 
standard model of guarded recursion, the topos of trees~\cite{first-steps}, which models types as presheaves over $\omega$, 
the ordered natural numbers. One important use case of guarded recursion is for denotational semantics of programming languages
with recursion~\cite{DBLP:journals/entcs/PaviottiMB15,DBLP:journals/mscs/MogelbergP19} also in combination
with other features such as probabilistic choice~\cite{StassenMZAB25} or non-determinism~\cite{TwoPowerDomains}. 
This programme is often referred to as \emph{synthetic guarded domain theory}.

Unfortunately, due to limited support for guarded recursion in proof assistants (see \autoref{sec:related} below for a detailed discussion),
none of the above mentioned applications have been formalised in proof assistants. One of the main challenges for doing so is that 
the language they are expressed in -- Guarded Type Theory~\cite{DBLP:conf/lics/BahrGM17} -- uses Fitch style~\cite{clouston2018fitch} for programming and reasoning about the $\tlater{}$ modality,
and such languages are not easily implemented as shallow embeddings in type theory, because type checking 
requires access to the context of expressions. 

This paper presents \grean, a framework for guarded recursion in Lean~4~\citep{DBLP:conf/cade/Moura021}, 
based on a deep embedding. 
Deep embeddings have a reputation for being expensive to work with, but
two design decisions make ours feasible while preserving expressivity. 
The first is the decision to not implement the full dependent type theory used in previous applications. Rather, 
we implement a simply typed guarded $\lambda$-calculus, with a higher-order logic (also with guarded recursion) over it. This has
the benefit that type checking never needs to normalize terms, and we do normalization only inside proofs. 
The second is that the typing and logical derivability are valued in Lean's definitionally proof irrelevant propositional universe $\Prop$, 
so that derivations are never unfolded or compared.
Taken together, these two aspects make quotation cheap: A \garlene program typechecked once, and the produced proof
is not unfolded when the program is referred to in other programs. 

\garlene is built in two layers. 
The first layer is the $\lambda$-calculus and logic, together with the tooling needed to work with a deep embedding: a typechecker for the  $\lambda$-calculus, and a proof mode with a set of tactics for the logic, which allows one to prove statements internal to our calculus.
The second layer is a model of both the calculus and the logic in the topos of trees, 
formalized in Lean using Mathlib~\citep{mathlib2020}.
The model serves two purposes: To prove the calculus and the logic sound, and to serve as a bridge between \grean\ and ordinary Lean. 
Theorems proved in \grean\ can be exported to Lean by interpretation into the model, and theorems proved about the model can be 
imported into \grean. To facilitate this bridge between \grean\ and Lean, our calculus and logic are both expandable: Any presheaf in the
topos of trees can be used as a type in \grean, and likewise any term and proof. 
One example of this is the way we include
guarded recursive types in the language: The model has a universe $\Usem$ of semantic types, and any map $\tlater\Usem \to \Usem$
in the model has a unique fixed point, which can be included into \grean\ as a type. Likewise the terms for folding and unfolding
the guarded recursive type, as well as the proofs that they constitute an isomorphism are imported from the model.

We illustrate the use of \garlene by two case studies.
The first is the guarded delay monad $D$ defined as the guarded recursive type satisfying 
$DA \iso A + \tlater (DA)$. This is a guarded variant of Capretta's delay monad~\cite{DBLP:journals/lmcs/Capretta05}
and is used in synthetic guarded domain theory as a monad for recursion. We show that this is a monad and 
that $DA$ is the free delay algebra on $A$, where a delay algebra is a type $B$ with an operation $\tlater B \to B$.
The second is a programming-language case study, a sound and adequate denotational semantics for a 
typed $\lambda$-calculus with fixpoints into a guarded universal domain satisfying equation \eqref{eq:untyped:lambda:model} above.
The resulting proof scripts match ordinary Lean tactic proofs in size and shape; \Cref{fig:teaser-proof} shows a representative one.

In summary, our contributions are:
\begin{enumerate}
\item an implementation in Lean of a Fitch-style $\lambda$-calculus with guarded recursion, 
and a higher-order logic over it. The calculus is a restriction of Guarded Type Theory~\cite{DBLP:conf/lics/BahrGM17} to 
simple types and a single global clock. The logic is a 
new adaptation of guarded type theory. 
\item a Kripke-style proof mode built from Lean tactics, in which guarded proofs are ordinary Lean tactic proofs 
\item two case studies: the delay monad, and a sound semantics for a typed $\lambda$-calculus with fixpoints 
in a guarded universal domain, proved adequate for the well-typed fragment of the language \item a set of implementation techniques and design decisions for working with deeply embedded calculi in Lean.
\end{enumerate}

The implementation is written in Lean~4; all code shown in the paper is taken verbatim from it.

\paragraph{Overview.}
\Cref{sec:related} discusses related work. \Cref{sec:overview} gives an informal overview of \garlene 
using the example of the guarded delay monad.
The following sections then treat the calculus (\Cref{sec:lang}), the logic (\Cref{sec:logic}), the proof mode (\Cref{sec:proofmode}) 
and the case studies (\Cref{sec:eval}). Finally, \Cref{sec:conclusion} summarizes this work and provides some notes on the future directions for this project.

 \section{Related work}
\label{sec:related}

\paragraph{Implementations of guarded type theory} Guarded Cubical Agda~\citep{DBLP:journals/jlap/VeltriV23} is an extension of Cubical 
Agda~\cite{DBLP:journals/pacmpl/VezzosiM019} with guarded recursion. It implements Clocked Cubical Type Theory~\cite{CubicalCloTT}, which combines guarded
recursion with cubical type theory~\cite{CTT}. One of the benefits of this combination is that cubical type theory
allows for extensionality principles to compute. In \grean, these extensionality principles are stated
using the proof irrelevant universe of propositions, and so do not need to compute. The modality $\tlater{}$, as well as the notion of ticks used for Fitch-style programming with $\tlater{}$ are native to 
Guarded Cubical Agda, and the Fitch-style typing rules are built into the type checker. This means that 
there is no distinction between the guarded type theory and native Cubical Agda like the distinction
between \grean\ and Lean in this work. Instead, Guarded Cubical Agda implements a version of guarded
recursion where $\tlater{}$ is indexed by clocks. In this setting, working in a context with no free clock variables
can be thought of as working in ordinary Cubical Agda, and working in a context with one free clock variable
corresponds to working in a cubical version of the topos of trees. One can pass between the two by 
weakening in one direction, and by quantification over clocks in the other. Guarded Cubical Agda
has been used for a few verification projects~\cite{DBLP:journals/pacmpl/GiovanniniDN25}, 
but is in practice a stand-alone proof assistant that is no longer maintained. 

Sikkel and BiSikkel~\citep{DBLP:journals/corr/abs-2207-00843,DBLP:journals/pacmpl/CeulemansND25} implement multimode type theories~\cite{DBLP:journals/lmcs/GratzerKNB21}, including guarded type theories, as deep embeddings in Agda together with a semantic interpretation, and are in this respect the closest to our work.
Our approach differs in two ways. First, BiSikkel allows only for direct manipulation of deeply embedded syntax, whereas \garlene includes an elaboration from user syntax, and terms can moreover be constructed interactively by tactics. 
This gives a development experience closer to that of ordinary Lean proofs.
Second, by using proof-irrelevant propositions we can carry typing and provability derivations in terms without ever inspecting them. 
This represents a major improvement in efficiency and allows for larger applications.

\paragraph{Logics with a later modality}
Iris~\citep{DBLP:journals/jfp/JungKJBBD18} provides the later modality, L\"ob induction (the logical correspondent
of $\fix$), and guarded recursive types and predicates. 
The Iris framework and MoSeL~\citep{DBLP:conf/popl/KrebbersTB17,DBLP:journals/pacmpl/KrebbersJ0TKTCD18} have become one of the biggest logics embedded in a proof assistant.
This is the closest existing practice to ours, and our proof mode follows the same general approach: tactics manipulate a reflected sequent, and the kernel checks the result.
One difference lies in how the language is constructed: Iris uses a shallow embedding of terms manipulating the denotational
model. While allowing a more direct implementation, this means that the Fitch-style modal languages that we use for programming and reasoning are unavailable.
In addition to that, the denotational models are different: Iris uses a category of (complete) bounded bisected ultrametrics, which embeds fully faithfully into the topos of trees. 
Although we do not use objects that lie outside this category in this work, we may do so in future extensions. 
For example, comprehension types $\{ x: A \mid \phi(x)\}$ generally do not live in the subcategory, but do live in the topos of trees. 
In future extensions, we may also want to change the indexing to ordinals larger than $\omega$
for programming with finite powersets, or reasoning about termination. Although a similar extension
exists for Iris~\cite{TransfiniteIris}, 
it is unknown how to model general guarded recursive types in these, as is possible
in extended versions of the topos of trees~\cite{mogelberg:beyondOmega}.

\paragraph{Frameworks for synthetic mathematics in Lean}
SynthLean~\citep{10.1145/3779031.3779087} is a framework for synthetic reasoning in Lean with
interpretations into natural models of dependent type theory. It has the same overall architecture as our system: surface syntax is elaborated into a deep embedding and then interpreted in a semantic model. SynthLean could in principle be instantiated with the topos of trees model,
but offers no support for Fitch-style programming with modalities as it relies on a single context.
 \section{Overview of \grean}
\label{sec:overview}

In this section we provide a short demo of \grean. We follow one
example, namely the guarded delay monad, through the whole system:
from its type declaration and programs (\Cref{fig:teaser-defs}),
through proofs about these programs (\Cref{fig:teaser-proof}), to an
extracted Lean theorem. To this end, we informally introduce \grean as
we go along, deferring the formal presentation of the language and its
logic until \Cref{sec:lang} and \Cref{sec:logic}, respectively.
Everything shown below is verbatim from the Lean implementation.

\begin{figure}
\begin{lstlisting}
gtype Delay (A : TYPE) := ν X. [A] ⊕ ▸X

gdef Delay.map (A B : TYPE) :
    (A → B) → [Delay A] → [Delay B] :=
  fix μ. λ f. λ d.
    [Delay.MK B]ₛ (case ([Delay.PROJ A]ₛ d)
      (λ a. inl (f a))
      (λ w. inr (delay (((adv 1 μ) f) (adv 1 w)))))
\end{lstlisting}
\caption{The guarded delay type and its functor action.}
\label{fig:teaser-defs}
\end{figure}

\subsection{A guarded type}
\label{sec:teaser-type}

We start with the type declaration in~\Cref{fig:teaser-defs}.
The command \lstinline|gtype| declares a guarded recursive type.
Here \lstinline|Delay| is parametrized by \lstinline|A : TYPE|, where
\lstinline|TYPE| is the type of \grean types, and is defined as a fixpoint: \lstinline|ν X| binds the recursion variable \lstinline|X|, and \lstinline|[A] ⊕ ▸X| says that a value of type \lstinline|Delay A| is either a value of type \lstinline|A|, available now, or another \lstinline|Delay A|, available only after taking a step.
The latter is expressed by the later modality \lstinline|▸|. A value of type \lstinline|▸T| is a value of type \lstinline|T| that becomes available one step in the future.
Unrolling the definition, an element of type \lstinline|Delay A| is a value of type \lstinline|A| after some finite number of steps, or a computation that never returns a value; this is the guarded version~\citep{DBLP:journals/entcs/PaviottiMB15} of Capretta's delay monad~\citep{DBLP:journals/lmcs/Capretta05}.
Note that \lstinline|X| occurs only under \lstinline|▸|.
This is what makes the fixpoint well-defined, and it is the only requirement: there is no positivity requirement.

Two auxiliary notations appear already in this example and are used later.
First, \lstinline|A| is a Lean variable, and square brackets, as in \lstinline|[A]| or \lstinline|[Delay A]|, splice a Lean expression denoting a type into the syntax of \grean.
Second, \lstinline|[c]ₛ| refers to a previously defined guarded program \lstinline|c| of \grean. We call this quotation (cf.\ \Cref{sec:quotation}).

Besides the type itself, the command generates two constants for folding and unfolding the fixpoint,
\begin{lstlisting}
Delay.MK A   : [A] ⊕ ▸[Delay A] → [Delay A]
Delay.PROJ A : [Delay A] → [A] ⊕ ▸[Delay A]
\end{lstlisting}
together with two equations stating that they are mutually inverse, which is witnessed by the generated \lstinline|Delay.MK_PROJ| and \lstinline|Delay.PROJ_MK|.
These two constants and two equations are the only interface to the type, and the fixpoint is not unfolded definitionally.
\begin{figure}
\begin{lstlisting}[escapeinside={(*@}{@*)}]
gtheorem Delay.map_id (A : TYPE) :
    ∀ d : [Delay A].
      (([Delay.map A A]ₛ (λ x : A. x)) d = d) := by
    glöb IH
    gintro d
    gunfold Delay.map
    gfix
    gcases ([Delay.PROJ A]ₛ d) with (⟨v, h⟩ | ⟨v, h⟩)
    · grewrite h
      gsimpl
      grewrite ← h
      gapply (Delay.MK_PROJ A)
    · grewrite h
      gsimpl
      gassert Htlf of ((delay _) = delay (adv 1 v))
      · gmono IH as G
        (*@\hlline{gapply G}@*)
      grewrite Htlf
      grewrite (Delay.delay_eta A) v
      grewrite ← h
      gapply (Delay.MK_PROJ A)
\end{lstlisting}
\caption{The functor identity law for \tac{Delay.map}. The two top-level bullet blocks are the \tac{inl} and the \tac{inr} case of the scrutinee. The highlighted line is the point where the proof state shown in \Cref{sec:teaser-proofs} is taken.}
\label{fig:teaser-proof}
\end{figure}

\subsection{Guarded terms}
\label{sec:teaser-terms}

Programs are written with the command \lstinline|gdef|, which typechecks the body once, at definition time, and makes it available as a constant in \grean.

Next, we define a function \lstinline|Delay.map|, which is our first guarded recursive definition.
\lstinline|fix μ| binds the name of the recursive function \lstinline|μ|, but at type \lstinline|▸((A → B) → [Delay A] → [Delay B])|: the function being defined is available only one step in the future.
The body proceeds by case analysis on \lstinline|[Delay.PROJ A]ₛ d|.
If the value is available now, we call \lstinline|f| immediately.
Otherwise we have \lstinline|w : ▸[Delay A]|, and we have to make a recursive call on it, which can only be done a step later.
This is where the two remaining primitives, \lstinline|delay| and \lstinline|adv|, come into play. 
They are the Fitch-style introduction and elimination forms for the later modality.
\lstinline|delay e| constructs a value of type \lstinline|▸T| from \lstinline|e : T|.
Inside its body we are one step in the future, and only there may \lstinline|adv| be applied to values of type \lstinline|▸T| from outside, such as \lstinline|μ| and \lstinline|w|, to get hold of the values they promise.
The numeral in \lstinline|adv 1| counts how many \lstinline|delay|s we look out through; in the examples it is always \lstinline|1|.
Since \lstinline|μ| can be used only under a \lstinline|delay|, every recursive call is made after a step, and \lstinline|Delay.map| is productive by construction; there is no separate termination or guardedness check.
\Cref{sec:ticks} makes this precise using a Kripke-style \emph{stack of contexts}, where each \lstinline|delay| opens a new frame and \lstinline|adv n| reaches back \lstinline|n| frames.

\subsection{Proofs: normalization and tactics}
\label{sec:teaser-proofs}

Properties of programs are stated with \lstinline|gtheorem| and proven in a higher-order logic over the calculus, which we present in~\Cref{sec:logic}.
\Cref{fig:teaser-proof} shows the functor identity law for \lstinline|Delay.map|.
The statement is again written in the syntax of \grean.
The proof is an ordinary Lean \lstinline|by|-block. Every step of the proof script uses a tactic provided by \grean for manipulating proof goals in \grean's logic. 
We also get goals between steps, error messages, and hover information, which is delaborated to hide implementation details and look more readable than a raw goal representation. The names of 
\grean's tactics are prefixed with \lstinline|g|, and most of them behave like their Lean counterparts: \lstinline|gintro|, \lstinline|gcases|, \lstinline|grewrite|, \lstinline|gapply|.
The tactics are described in \Cref{sec:proofmode}.

The proof is by L\"ob induction, initiated by the tactic \lstinline|glöb IH|. Writing \lstinline|P| for the current goal, \lstinline|glöb IH| 
introduces an assumption 
\lstinline|IH : lift (delay P)|, i.e.\ \lstinline|P| one step later, which we may use to prove 
\lstinline|P|. 
We will write the composition of \lstinline|lift| and \lstinline|delay| as $\later$, and this is a propositional
version of \lstinline|▸|. After unfolding the definition of \lstinline|Delay.map| using \lstinline|gfix|, the proof branches on 
the value of \lstinline|[Delay.PROJ A]ₛ d|. The interesting case  
is \lstinline|[Delay.PROJ A]ₛ d = inr v| for some \lstinline|v : ▸ [Delay A]|. In this case \lstinline|[Delay.map A A]ₛ (λ x : A. x)) d| unfolds
to \lstinline|[Delay.MK A]ₛ (inr (delay (... (adv 1 v))))|, where the elided part is the unfolded recursive call. So it suffices 
to prove that \lstinline|delay (... (adv 1 v)) = v|, an equality between terms of delayed type. By $\eta$-unfolding of the right hand side,
this is the goal \lstinline|Htlf|. To prove such an equality between delayed data, it suffices to prove that the data delivered by each side
in the next step are equal. The tactic \lstinline|gmono IH as G| makes this move by moving into the future and making the induction 
hypothesis without $\later$ available as \lstinline|G|. 
Lean displays the goal at the highlighted line of \Cref{fig:teaser-proof} as follows (we omit the unfolded body of \lstinline|Delay.map|).

\noindent\begin{minipage}{\columnwidth}
\begin{lstlisting}
d : Delay A
v : ▸ [Delay A]
IH : lift (delay (∀ d. ([map A A]ₛ (λ x. x)) d = d))
h : [PROJ A]ₛ d = inr v
─────▷─────
G : ∀ d. ([map A A]ₛ (λ x. x)) d = d
⊢ ((fix μ. ...) (λ x. x)) (adv 1 v) = adv 1 v
\end{lstlisting}
\end{minipage}

\noindent The horizontal line separates the two time frames: the hypotheses above it were introduced before the step, \lstinline|G| after it.
The goal then follows by application of \lstinline|G|.
The rest of the proof is a chain of rewritings: with \lstinline|Htlf|, with the $\eta$-law for \lstinline|delay|, and back with \lstinline|h|, after which the generated equation \lstinline|Delay.MK_PROJ| closes the goal, similar to the \lstinline|inl| branch.

\subsection{Denotation}
\label{sec:teaser-denotation}

So far, all definitions and proofs happened inside \grean.
What connects \grean to the rest of Lean is the model: Types of \grean are interpreted as objects of the topos of trees $\topos$, programs as morphisms and propositions as morphisms into the subobject classifier (\Cref{sec:semantics}).
Soundness (\Cref{sec:soundness}) then says that every theorem proven in the proof mode holds in the model.
For \lstinline|Delay.map_id| this gives an ordinary Lean theorem about morphisms of $\topos$:
\begin{lstlisting}
theorem Delay.map_id_denotation (A : TYPE.{u}) :
    denoteHom (box([Delay.map A A]ₛ [idfun A]ₛ))
      = 𝟙 (⟦Delay A⟧ₜ : ℐ.{u})
\end{lstlisting}
Here \lstinline|⟦Delay A⟧ₜ| is the presheaf interpreting the type \lstinline|Delay A|, \lstinline|denoteHom| interprets a closed program of function type as a morphism between the interpretations of its domain and codomain, \lstinline|box(...)| is the anonymous form of quotation, which packages a closed \grean term, and \lstinline|idfun A| is the identity function defined with \lstinline|gdef|.
The statement mentions no \grean proposition, only its interpretation.
In the same way, the monad laws of \lstinline|Delay|, all proven in the proof mode, assemble into an instance \lstinline|DELAY.monad : CategoryTheory.Monad ℐ| of Mathlib's monads on the topos of trees (\Cref{sec:delay}).

The model also works in the other direction: A morphism of $\topos$ can be used as a constant in \grean (using \lstinline|ax|), and two programs with equal interpretations are equal in the logic.
This is how, for instance, the Mathlib morphisms in the statement of the monad laws enter \grean.
For closed propositions there is, moreover, a small extraction interface (\Cref{sec:extraction}).
A proposition is \emph{valid} if its interpretation holds in the model; every theorem of the proof mode is valid by soundness; a theorem \lstinline|P → Q| and a valid \lstinline|P| give a valid \lstinline|Q|; a valid $\later P$ gives a valid $P$ (validity quantifies over all steps at once, so a $\later$ may be stripped, even though $\later P \to P$ is not a theorem of the logic); and a valid \lstinline|pure φ|, for a Lean proposition \lstinline|φ|, gives \lstinline|φ| itself.
The last two rules are how results leave the guarded model: the steps taken during a proof are discarded, and \Cref{sec:adequacy} uses this to turn a guarded adequacy proof into a statement about an operational semantics that mentions neither \grean nor the topos and is stated using Lean's proposition universe.

 \section{The \garlene\ core calculus}
\label{sec:lang}

In this section, we give a precise account of \grean's core calculus, 
a simply typed $\lambda$-calculus with products,
sums, the later modality, and a type $\Omega$ of propositions. In
addition, \grean has two type formers $\Delta$ and $\syn{ax}$ for
referring to objects from the surrounding Lean context:
\[
  \text{types} \quad  \type, \sigma ::= \Delta\,T \mid \tprod{\type}{\sigma} \mid \type \mathbin{\syn{\oplus}} \sigma \mid \tarr{\type}{\sigma} \mid \tlater{\type} \mid \Omega \mid \syn{ax}\;I
\]
$\Delta\,T$ embeds a Lean type $T$ as a discrete \grean type, and
$\syn{ax}\;I$ embeds a presheaf $I$ of the model as a \grean type. 
Expressions include both terms and propositions:
  \[
    \arraycolsep=2pt
    \begin{array}{llcl}
  \text{expr.}\; & e, \Phi & ::=\; & \delta(a) \mid e \odot e' \mid \syn{pure}\;e \mid \syn{var}\;(p,q)
   \mid \syn{lam}_{\type}\;e \\
  &&\mid&\syn{app}_{\type}\;e\;e' \mid \vdelay{e} \mid \eadv{n}{e} \mid \syn{fix}_{\tlater{\type}}\;e \mid \vpair{e}{e'} \\
  &&\mid& \syn{proj}\;e\;d \mid \syn{inl}_{\type}\;e \mid \syn{inr}_{\type}\;e \mid \syn{case}_{\type,\sigma}\;e\;f\;g  \\
  &&\mid& \Phi \wedge \Phi' \mid \Phi \vee \Phi' \mid \Phi \to \Phi' \mid \forall_{\type}\,\Phi \mid \exists_{\type}\,\Phi  \\
  &&\mid& \elift{e} \mid \top \mid \bot \mid e =_{\type} e' \mid \syn{ax}_{\type}\;f
    \end{array}
\]
The syntax uses de Bruijn indices, which are pairs of the form
$\syn{var}\,(p,q)$ to account for the Kripke-style typing of \grean,
which we describe in \Cref{sec:ticks}. Four of the expression formers
refer to Lean. The expression $\delta(a)$ embeds a Lean value $a \col
T$ at the discrete type $\Delta\,T$, which together with $\odot :
\Delta\,(S \to T) \to \Delta\,S \to \Delta\,T$ gives discrete types an
applicative structure. The expression $\syn{pure}\;e$ turns an
embedded Lean proposition into a \grean proposition, and
$\syn{ax}_{\type}\;f$ uses a global element $f \from \One \longto
\sem{\type}$ of the model as a closed constant of type $\type$. The
remaining formers are standard, with the exception of $\syn{delay}$,
$\eadv{n}{}$ and $\syn{fix}$, which interact with the context in a way
that we describe in \Cref{sec:ticks}. $\elift{e}$ turns a delayed
proposition $e \col \tlater{\Omega}$ into a proposition now, allowing
us to obtain a propositional version of the later modality
(\Cref{sec:logic}).

\begin{figure*}
\begin{figrules}
\begin{mathpar}
\inferrule*[lab=embed]
  {|\vec{\Gamma}| > 0}
  {\typed{\vec{\Gamma}}{\delta(a)}{\Delta\,T}}
\and
\inferrule*[lab=embed-apply]
  {\typed{\vec{\Gamma}}{f}{\Delta(T \to S)} \\ \typed{\vec{\Gamma}}{x}{\Delta\,T}}
  {\typed{\vec{\Gamma}}{f \odot x}{\Delta\,S}}
\and
\inferrule*[lab=pure]
  {\typed{\vec{\Gamma}}{e}{\Delta\,\Prop}}
  {\typed{\vec{\Gamma}}{\syn{pure}\;e}{\Omega}}
\and
\inferrule*[lab=ax]
  {|\vec{\Gamma}| > 0 \\ f \from \One \longto \sem{\type}}
  {\typed{\vec{\Gamma}}{\syn{ax}_{\type}\;f}{\type}}
\and
\inferrule*[lab=lift]
  {\typed{\vec{\Gamma}}{e}{\tlater{\Omega}}}
  {\typed{\vec{\Gamma}}{\elift{e}}{\Omega}}
\\
\inferrule*[lab=var]
  {\Gamma_p = \type_0,\dots,\type_q,\Gamma}
  {\typed{\Gamma_0;\dots;\Gamma_p;\vec{\Gamma}}{\syn{var}\;(p,q)}{\type_q}}
\and
\inferrule*[lab=lam]
  {\typed{(\type, \Gamma) ; \vec{\Gamma}}{e}{\sigma}}
  {\typed{\Gamma ; \vec{\Gamma}}{\syn{lam}_{\type}\;e}{\tarr{\type}{\sigma}}}
\and
\inferrule*[lab=delay]
  {|\vec{\Gamma}| > 0 \\ \typed{\cdot\; ; \vec{\Gamma}}{e}{\type}}
  {\typed{\vec{\Gamma}}{\vdelay{e}}{\tlater{\type}}}
\and
\inferrule*[lab=adv]
  {n > 0 \\ \typed{\vec{\Gamma}}{e}{\tlater{\type}}}
  {\typed{\Gamma_1;\dots;\Gamma_n;\vec{\Gamma}}{\eadv{n}{e}}{\type}}
\and
\inferrule*[lab=fix]
  {\typed{(\tlater{\type} , \Gamma) ; \vec{\Gamma}}{e}{\type}}
  {\typed{\Gamma ; \vec{\Gamma}}{\syn{fix}_{\tlater{\type}}\;e}{\type}}
\end{mathpar}
\end{figrules}
\caption{Typing rules, fragment: the formers that refer to Lean and the formers that interact with the frame structure of contexts.
A context $\vec{\Gamma}$ is a stack of frames $\Gamma$ (\Cref{sec:ticks}).
The rules for application, products, sums, connectives, quantifiers, equality, $\top$ and $\bot$ are omitted.
In \textsc{ax}, $f \from \One \longto \sem{\type}$ is a global element of the model.
Well-typedness side conditions are omitted except where they matter.}
\label{fig:typing}
\end{figure*}

\subsection{Type system}
\label{sec:ticks}
\Cref{fig:typing} shows the fragment of the typing rules that are
non-standard. \grean uses a Kripke-style type
system~\citep{DBLP:journals/jacm/DaviesP01,DBLP:journals/jfp/HuJP23} whose contexts are not
lists of types but rather \emph{stacks of frames} (or alternatively, \emph{stacks of contexts}). 
A stack $\vec\Gamma$ is a list $\Gamma_0;\dots;\Gamma_n$ of frames $\Gamma_i$,
and in turn a frame $\Gamma$ is a list $\tau_0,\dots,\tau_n$ of types.
The Kripke-style type system is a presentation of the Fitch-style type system of Clocked Type
Theory~\citep{DBLP:conf/lics/BahrGM17} on which \grean is based. The
``;'' separators between frames correspond to the ticks in Clocked Type Theory, or the padlocks in
the Fitch-style type system of \citet{clouston2018fitch}. 
They represent the 'steps' of \autoref{sec:overview}. From here on we use the terminology
'tick' rather than 'step'.

Since we use de Bruijn variables, we append both the stacks and individual contexts from the left for consistency.
Consequently, de Bruijn variables are pairs $(p,q)$, where $p$ refers
to the frame $\Gamma_p$ in the stack $\vec\Gamma$, and $q$ refers to
the entry $\tau_q$ in that frame $\Gamma_p$. We write $\vec{\Gamma}$,
$\vec{\Delta}$ for contexts and $\Gamma$, $\Delta$, $\Psi$ for single
frames. The context of any derivable judgment has at least one frame,
and the closed context is the single empty frame denoted $\cdot$. This
is the reason for the side condition $|\vec{\Gamma}| > 0$ in several
rules. The expression $\vdelay{e}$
is typechecked with a new empty frame pushed onto the context, which
means that inside a $\syn{delay}$ we are one tick in the future, and
the variables of the outer frames are one tick in the past. The
expression $\eadv{n}{e}$ typechecks $e \col \tlater{\type}$ in the
context with the innermost $n$ frames removed and produces a value of
type $\type$. It reaches back $n$ ticks for a delayed value and uses
it now. Expressions with binders, such as $\syn{lam}$ or
$\syn{\forall}$, introduce a variable in the innermost frame. Finally,
$\syn{fix}\;e$ binds the recursive occurrence in the innermost frame
at type $\tlater{\type}$. Consequently, the recursive occurrence can
be used only at type $\tlater{\type}$, typically by applying
$\syn{adv}$ to it under a $\syn{delay}$, as shown
in~\Cref{sec:overview}.

Fitch-style programming with $\syn{delay}$ and $\syn{adv}$ 
gives an intuitive way of programming with the $\tlater{}$ modality. 
The cost of this is that the theory of substitutions more complex, 
as we describe in the following section.

\subsection{Weakening and substitution}
\label{sec:weakening}

Renamings and substitutions are defined by the following grammars:
\[
    \arraycolsep=2pt
    \begin{array}{llcl}
\text{renamings}\; & \ren &::=\; &\rid \mid \ren \rcomp \ren' \mid \rwkv{\ren} \mid \rliftv{\ren} \mid \rwkf{\ren}{n} \mid \rliftf{\ren}\\
  \text{frame subst.}\; & \sbst &::=\; & \snil \mid e \scons \sbst \\
  \text{stacked subst.}\; & \ssbst &::=\; & \sbst \mid \sbst \sframe{n} \ssbst
    \end{array}
 \]
Apart from identity and composition renamings, there are two
dimensions, with two operations in each. Within a frame, one can
weaken past an entry ($\rwkv{}$) or lift a renaming under a binder
($\rliftv{}$); across frames, one can weaken past $n$ frames
($\rwkf{}{n}$) or lift a renaming under a $\syn{delay}$ ($\rliftf{}$).
Substitutions are structured in the same way and inherit the notation.
A frame substitution $\sbst$ populates a single frame, while a stacked
substitution $\ssbst$ populates a stack of frames.
When a frame substitution is well-typed, $\subty{\sbst}{\vec{\Gamma}}{\Delta}$, 
we can upgrade it to a stacked substitution to a singleton context: ${\subty{\sbst}{\vec{\Gamma}}{[\Delta]}}$.
Lifting under a binder is not a primitive substitution but a derived construction, 
$\sliftv{\sbst} = \syn{var}\,(0,0) \scons \act{\sbst}{\rwkv{\rid}}$, 
where the renaming acts on $\sbst$ entrywise; 
$\sliftv{\ssbst}$ applies this to the innermost frame of $\ssbst$ 
and leaves the remaining frames unchanged.
\Cref{fig:ren}
defines well-formedness judgments for renamings and substitutions, as
well as their respective actions $\act{e}{\ren}$ and $\bnd{e}{\ssbst}$
on a term $e$.

The substitution action $\bnd{e}{\ssbst}$ will appear in the
provability and equality judgments of the logic in \Cref{sec:logic},
but only in the special case where $\ssbst = e'\scons \snil$, which we
write as $\bnd{e}{e'}$ instead of $\bnd{e}{e'\scons\snil}$. Most
clauses of the renaming and substitution actions are congruences and
\Cref{fig:ren} shows the cases that cross a tick or a variable binder.
Binders and $\syn{delay}$ lift the renaming into the new slot or
frame. For $\eadv{n}{e}$, the number $n$ is replaced by the number of
frames onto which the renaming maps the $n$ frames that $\syn{adv}$
crosses, which we call the offset $\offset{\ren}{n}$, and the action
continues on $e$ with the part of the renaming that lies beyond those
frames, which we call the cut $\cut{\ren}{n}$.\footnote{Our cut and
offset are the truncation $\sigma \mid n$ and the truncation offset
$\mathcal{O}(\sigma, n)$ of~\citet{DBLP:journals/jfp/HuJP23}; what
they call the modal offset is our $n$ in $\eadv{n}{}$.}
Offset and cut extend to substitutions. 
$\offset{\ssbst}{n}$ is the sum of the frame counts $m$ of the first 
$n$ components $\sframe{m}$ of $\ssbst$, and $\cut{\ssbst}{n}$ is 
$\ssbst$ with those components removed.
Stacked substitutions are the K-substitutions
of~\citet{DBLP:journals/jfp/HuJP23} in a simply typed setting: the
frame $\sframe{n}$ is their modal extension $\sigma;\Uparrow^{n}$, and
the clause for $\syn{adv}$ in \Cref{fig:ren} is their rule for pushing
a substitution under $\syn{unbox}$. Unlike their calculus, where
K-substitutions are part of the syntax and typing and conversion are
defined mutually, our renamings and substitutions are functions on raw
terms, and preservation of typing is a lemma rather than a rule.

Renamings and substitutions enjoy the expected properties.
Typing is preserved: from $\typed{\vec{\Delta}}{e}{\type}$ and $\renty{\ren}{\vec{\Gamma}}{\vec{\Delta}}$ we obtain $\typed{\vec{\Gamma}}{\act{e}{\ren}}{\type}$, which every rule that moves a term between contexts relies on.
Renamings and substitutions are trivial on closed terms: a closed, well-typed term is a fixed point of every well-typed renaming and substitution, so a quoted constant may be placed at any offset, which we exploit for quotation below.
Index arithmetic occurs only in the two clauses for $\syn{adv}$, and every other clause passes the renaming either unchanged or lifted.
So, renamings re-index variables, adjust the numbers in $\eadv{n}{}$, and leave every other former unchanged.
Renamings also satisfy the expected identity and composition laws, so the administrative renamings produced by tactics compose and cancel by simplification with these laws, and renamings do not appear in user-displayed goals.
Cut and offset satisfy the coherence laws of the truncations similar to~\citep{DBLP:journals/jfp/HuJP23}: $\cut{\ren}{a{+}b} = \cut{(\cut{\ren}{a})}{b}$ and $\offset{\ren}{a} + \offset{(\cut{\ren}{a})}{b} = \offset{\ren}{a{+}b}$.

\begin{figure*}
\begin{figrules}
\begin{mathpar}
\inferrule*[lab=id]
  {|\vec{\Gamma}| > 0}
  {\renty{\rid}{\vec{\Gamma}}{\vec{\Gamma}}}
\and
\inferrule*[lab=comp]
  {\renty{\ren_1}{\vec{\Delta}}{\vec{\Psi}} \\ \renty{\ren_2}{\vec{\Gamma}}{\vec{\Delta}}}
  {\renty{\ren_1 \rcomp \ren_2}{\vec{\Gamma}}{\vec{\Psi}}}
\and
\inferrule*[lab=local-weaken]
  {\renty{\ren}{\Gamma ; \vec{\Gamma}}{\Delta ; \vec{\Delta}}}
  {\renty{\rwkv{\ren}}{(\type , \Gamma) ; \vec{\Gamma}}{\Delta ; \vec{\Delta}}}
\\
\inferrule*[lab=cons]
  {\renty{\ren}{\Gamma ; \vec{\Gamma}}{\Delta; \vec{\Delta}}}
  {\renty{\rliftv{\ren}}{(\type , \Gamma) ; \vec{\Gamma}}{(\type , \Delta) ; \vec{\Delta}}}
\and
\inferrule*[lab=global-lift]
  {\renty{\ren}{\vec{\Gamma}}{\vec{\Delta}}}
  {\renty{\rliftf{\ren}}{\cdot\, ; \vec{\Gamma}}{\cdot\, ; \vec{\Delta}}}
\and
\inferrule*[lab=global-shift]
  {\renty{\ren}{\vec{\Gamma}}{\vec{\Delta}}}
  {\renty{\rwkf{\ren}{n}}{\Psi_1 ; \dots ; \Psi_n ; \vec{\Gamma}}{\vec{\Delta}}}
\\
\inferrule*[lab=snil]
  {|\vec{\Gamma}| > 0}
  {\subty{\snil}{\vec{\Gamma}}{[\,]}}
\and
\inferrule*[lab=scons]
  {\typed{\vec{\Gamma}}{e}{\type} \\ \subty{\sbst}{\vec{\Gamma}}{\Delta}}
  {\subty{e \scons \sbst}{\vec{\Gamma}}{\type , \Delta}}
\and
\inferrule*[lab=frame]
  {n > 0 \\ \subty{\ssbst}{\vec{\Gamma}}{\vec{\Delta}} \\ \subty{\sbst}{\Psi_1 ; \dots ; \Psi_n ; \vec{\Gamma}}{\Delta}}
  {\subty{\sbst \sframe{n} \ssbst}{\Psi_1 ; \dots ; \Psi_n ; \vec{\Gamma}}{\Delta ; \vec{\Delta}}}
\end{mathpar}

\medskip
\[
  \begin{array}{@{}r@{\;=\;}l@{\qquad}r@{\;=\;}l@{}}
    \act{(\vdelay{e})}{\ren} & \vdelay{\act{e}{\rliftf{\ren}}}
      & \bnd{(\vdelay{e})}{\ssbst} & \vdelay{\bnd{e}{(\snil \sframe{1} \ssbst)}} \\[2pt]
    \act{(\eadv{n}{e})}{\ren} & \eadv{\offset{\ren}{n}}{\act{e}{(\cut{\ren}{n})}}
      & \bnd{(\eadv{n}{e})}{\ssbst} & \eadv{\offset{\ssbst}{n}}{\bnd{e}{(\cut{\ssbst}{n})}} \\[2pt]
    \act{(\syn{fix}\;e)}{\ren} & \syn{fix}\;\act{e}{\rliftv{\ren}}
      & \bnd{(\syn{lam}_{\type}\;e)}{\ssbst} & \syn{lam}_{\type}\;\bnd{e}{\sliftv{\ssbst}}
  \end{array}
\]
\end{figrules}
\caption{Typing of renamings (top two rows) and of substitutions (third row), and representative clauses of the two actions: those that cross a tick, and one binder case for each.
In \textsc{global-shift} and \textsc{frame}, $n$ is the number of added frames $\Psi_1, \dots, \Psi_n$.
A frame substitution $\sbst$ and a stacked one $\ssbst$ share the judgment symbol; the metavariable shows which is meant, and in the $\snil$ and $\scons$ rules the right-hand side is a single frame.}
\label{fig:ren}
\end{figure*}

\phantomsection
\label{sec:quotation}

\paragraph{Typechecking and quotation.}
\grean's syntax is defined as an inductive type \lstinline|EXPR|, which
is extrinsically typed by a Lean predicate \lstinline|TYPED : CTX → EXPR → TYPE → Prop|. A \grean term defined
via \lstinline|gdef| is typechecked by the \grean implementation,
which produces a proof of \lstinline|TYPED [.] e τ|. Such a closed
\grean term $e$ of type $\tau$ is then packaged as a Lean value of type
\lstinline|SYNT τ|. Since \lstinline|TYPED| is a proposition, the
derivation is proof-irrelevant: derivations are never compared, and
two packages with the same term are equal. We use this trick to have a
mechanism to reuse previously defined \grean programs. More
concretely, when users refer to a previously defined program $c$, the
elaborator inserts it as $\quoat{c}{k}{m}$, which re-embeds $c \col
\SYNT\;\type$ into a context of $k$ frames whose outermost frame has
$m$ slots, as the weakening $\act{c}{\wkquote{k-1}{m}}$. This allows
tactics to re-index quotations freely and to compare them by their
offsets. Until the user explicitly unfolds $c$, all administrative
weakenings and substitutions produced by tactics simply re-index
quotations around $c$. In the surface syntax a quotation is written
\lstinline|[c]ₛ| for a Lean value \lstinline|c : SYNT τ|, and
\lstinline|box(e)|, its anonymous form, which quotes a closed term in
place.

\subsection{Guarded recursive types}
\label{sec:rectypes}

The command \lstinline|gtype| introduces a guarded recursive type \lstinline|N| from a type body \lstinline|F| in which the recursion variable occurs under $\tlater{}$, either directly as in
\begin{lstlisting}
gtype Delay (A : TYPE) := ν X. [A] ⊕ ▸X
\end{lstlisting}
or under further constructors as in an untyped-$\lambda$-model~\citep{DBLP:journals/entcs/PaviottiMB15}:
\begin{lstlisting}
gtype Dom := ν X. Δ Nat ⊕ Δ Unit ⊕ ▸X ⊕ ▸(X → X)
\end{lstlisting}
It generates the isomorphism as a pair of object-language constants
\begin{lstlisting}
N.MK   : SYNT (F[N] → N)
N.PROJ : SYNT (N → F[N])
\end{lstlisting}
with proved equations \lstinline|N.MK_PROJ| and \lstinline|N.PROJ_MK| making them mutually inverse.
The recursive type itself is included into the language via the semantics as described below. 
The body of a \lstinline|gtype| declaration is written in the same type syntax as the signatures of \lstinline|gdef|, 
restricted to ensure that the recursion variable occurs only under \lstinline|▸|. 
For the moment this restricted language contains the type constructors needed for the examples:
core calculus types \lstinline|[τ]|, discrete types \lstinline|Δ T|, products, sums, \lstinline|▸X|, and \lstinline|▸(F → G)|, 
where either side of the arrow may be the recursion variable or a nested body.
That is, the recursion variable may occur only under \lstinline|▸|, but it may occur on either side of an arrow.
We plan to expand this language in future work.

\phantomsection
\label{sec:universes}
\paragraph{Semantics of guarded recursive types.}
The semantic model of the \garlene\ calculus is a presheaf category, and so has a 
Hofmann--Streicher universe $\Usem$~\citep{hofmann-streicher-lift}, the presheaf whose component at stage $n$ consists of the small 
presheaves on the stages $0,\dots,n$. Guarded recursive types are modelled as fixed points of maps of the type 
$\tlater{\Usem} \to \Usem$ following~\citet{DBLP:conf/lics/BirkedalM13}. 
Such fixed points are global elements of $\Usem$ and these correspond to objects of the topos of trees. 

\begin{figure}
\begin{lstlisting}
UNIV.DISCRETE T     : SYNT UNIV
UNIV.CODE τ         : SYNT UNIV
UNIV.PROD, UNIV.SUM : SYNT (UNIV × UNIV → UNIV)
UNIV.LATER          : SYNT (▸ UNIV → UNIV)
UNIV.LARR           : SYNT (▸ UNIV × ▸ UNIV → UNIV)
\end{lstlisting}
\caption{Code formers.}
\label{fig:codes}
\end{figure}

To program with the universe we include it into the \garlene calculus as $\UNIV := \syn{ax}\;\Usem$, and add codes 
for type formers as shown in~\Cref{fig:codes}. 
A guarded recursive type is then an ordinary fixpoint at type $\UNIV$,
\[
  \mathtt{N.code} \;:=\; \quo{\vfix{X}{\,\widehat{F}[X]}} \;\col\; \SYNT\;\UNIV ,
\]
with $\widehat{F}$ the body rewritten with the code formers.
Each of the codes is modelled in the topos of trees as morphisms corresponding to the appropriate type 
constructors via global elements. This ensures
that the fixed points correspond to objects satisfying the appropriate type equation up to identity: $\sem{N} = \sem{\plug{F}{N}}$.

The code $\widehat{F}$ has type $\tlater\UNIV \to \UNIV$ essentially because the
two code formers that involve the later modality, $\syn{LATER}$ and $\syn{LARR}$, take \emph{delayed} codes as arguments.
The \lstinline{gtype} command checks its body against the grammar above when parsing; the generated code is then typechecked like any term, and since these formers consume delayed codes, the \textsc{fix} rule certifies independently that every occurrence of $X$ is guarded.

\subsection{Semantics}
\label{sec:semantics}

The model is the topos of trees $\topos$~\citep{first-steps}, the presheaves over $\omega$. Recall that the objects of $\topos$ are
natural number indexed sets, together with maps $X(n+1) \to X(n)$ for all $n$. 
Types are interpreted by objects $\sem{\type}$, contexts by objects $\sem{\vec{\Gamma}}$, and a partial interpretation
\[
  \mathtt{expr\_interp}\;\vec{\Gamma}\;e\;\type \col \mathtt{Part}\,(\sem{\vec{\Gamma}} \longto \sem{\type})
\]
interprets terms.
The interpretation is partial because typing is extrinsic: it is a function on raw syntax, and a separate lemma states that it is defined whenever $\typed{\vec{\Gamma}}{e}{\type}$.
The later modality is interpreted by the $\tlater{}$ functor defined as $\tlater X(1) = 1$, $\tlater X(n+1) = X(n)$, 
$\Delta T$ by the constant presheaf on $T$, and $\Omega$ by the subobject classifier.
The formers that refer to Lean are interpreted directly: $\delta(a)$ by the constant map at $a$, $\odot$ by the application map of the constant presheaf, $\syn{pure}$ and $\elift{}$ by maps into the subobject classifier, and $\syn{ax}_{\type}\;f$ by $f$ precomposed with the terminal map $\sem{\vec{\Gamma}} \longto \One$.
The frame structure of contexts is interpreted through the left adjoint $\blacktriangleleft$ of the later functor, the \emph{earlier} functor: a frame is the product of the interpretations of its types, and each frame boundary is one application of $\blacktriangleleft$, so $\sem{\Gamma ; \vec{\Gamma}} = \sem{\Gamma} \times \blacktriangleleft\,\sem{\vec{\Gamma}}$.
The former $\syn{delay}$ is then interpreted by transposition along $\syn{\blacktriangleleft} \dashv \tlater{}$, turning $\syn{\blacktriangleleft}\,\sem{\vec{\Gamma}} \longto \sem{\type}$ into $\sem{\vec{\Gamma}} \longto \tlater{\sem{\type}}$.
The former $\eadv{n}{}$ projects the context onto its outer frames and transposes in the other direction; for $n > 1$ the surplus boundaries are first collapsed by iterating $\mathsf{force} \col \syn{\blacktriangleleft} X \longto X$, the transpose of 
$\syn{next} \col X \longto \tlater{X}$.
The same transpose interprets the use of a variable across a frame boundary, as noted in \Cref{sec:ticks}.
The interpretation of $\syn{fix}$ is a morphism $X^{\tlater{X}} \longto X$ defined by recursion on the stage: at stage $0$ the object $\tlater{X}$ is a singleton, so the fixpoint is determined trivially, and at stage $n+1$ it is obtained from stage $n$.
 \section{The logic}
\label{sec:logic}

In this section we describe the logic of \grean, its proof rules, and
its proof mode implemented in Lean.

\subsection{Rules}
\label{sec:rules}
A \grean proposition $\Phi$ is a term of type $\Omega$. The provability
judgment $\gseq{\vec{\Gamma}}{\vec{\Psi}}{\Phi}$ references a
hypothesis context $\vec{\Psi}$, which, like the variable context
$\vec{\Gamma}$, is structured into a stack of frames. The stack
represents a separation of logical assumptions into time frames, 
like in the proof example of \Cref{sec:teaser-proofs}, and these 
time frames are synchronised with the time frames for variables
in rules. This means that every rule will maintain
the invariants that $|\vec{\Gamma}| = |\vec{\Psi}|$, and that
every hypothesis is a well-typed proposition in the context up to the frame 
where it was introduced. We write this as the following judgment:
\[
\Gamma_0;\dots;\Gamma_n \vdash \Psi_0;\dots;\Psi_n
  \;\eqdef\;
  \forall k.\ \forall \Phi \in \Psi_k.\ \typed{\Gamma_k;\dots;\Gamma_n}{\Phi}{\Omega},
\]
In particular, a 
predicate can only mention variables defined in or before its own
time frame. This synchronisation of contexts is automatic
in dependent type theories like Clocked Type Theory~\cite{DBLP:conf/lics/BahrGM17},
where both variables and logical assumptions live in the same context. Our logic
was inspired by Clocked Type Theory, but to the best of our knowledge, this
is the first Fitch-style logic over a Fitch-style type theory with separate contexts
for variables and hypotheses.

\Cref{fig:proves} shows the rules that are specific to our logic, with
the well-typedness side conditions omitted. A \grean theorem is
declared using the command \lstinline|gtheorem|, which elaborates to
an ordinary Lean theorem whose statement encodes $\gseq{\cdot}{\cdot}{\Phi}$, i.e.\
$\Phi$ is a closed term of type $\Omega$ that is provable in the empty
hypothesis context.

\begin{figure*}
\begin{figrules}
\begin{mathpar}
\inferrule*[lab=asm]
  {\Phi \in \Psi_n}
  {\gseq{\vec{\Gamma}}{\Psi_0;\dots;\Psi_n;\vec{\Psi}}{\act{\Phi}{\wkf{n}}}}
\and
\inferrule*[lab=forall-intro]
  {\gseq{(\type , \Gamma) ; \frames}{\act{\vec{\Psi}}{\wkv}}{\Phi}}
  {\gseq{\Gamma ; \frames}{\vec{\Psi}}{\forall_{\type}\,\Phi}}
\and
\inferrule*[lab=forall-intro-points]
  {\forall a \col T.\, \gseq{\vec{\Gamma}}{\vec{\Psi}}{\bnd{\Phi}{\delta(a)}}}
  {\gseq{\vec{\Gamma}}{\vec{\Psi}}{\forall_{\Delta\,T}\,\Phi}}
\and
\inferrule*[lab=lift-intro]
  {\gseq{\cdot ; \vec{\Gamma}}{\cdot ; \vec{\Psi}}{\Phi}}
  {\gseq{\vec{\Gamma}}{\vec{\Psi}}{\later\Phi}}
\and
\inferrule*[lab=later-mono]
  {\gseq{\vec{\Gamma}}{\vec{\Psi}}{\later P} \\ \gseq{\cdot; \vec{\Gamma}}{P ; \vec{\Psi}}{Q}}
  {\gseq{\vec{\Gamma}}{\vec{\Psi}}{\later Q}}
\and
\inferrule*[lab=delay-eq]
  {\gseq{\vec{\Gamma}}{\vec{\Psi}}{\later(e_1 =_{\type} e_2)}}
  {\gseq{\vec{\Gamma}}{\vec{\Psi}}{\vdelay{e_1} =_{\tlater{\type}} \vdelay{e_2}}}
\and
\inferrule*[lab=loeb-ind]
  {\gseq{\vec{\Gamma}}{(\later(\act{\Phi}{\wkf{1}}) , \Psi) ; \vec{\Psi}}{\Phi}}
  {\gseq{\vec{\Gamma}}{\Psi ; \vec{\Psi}}{\Phi}}
\and
\inferrule*[lab=eq-def]
  {\typed{\vec{\Gamma}}{e_1 \equiv e_2}{\type}}
  {\gseq{\vec{\Gamma}}{\vec{\Psi}}{e_1 =_{\type} e_2}}
\and
\inferrule*[lab=eq-elim]
  {\gseq{\frames}{\vec{\Psi}}{e_1 =_{\type} e_2} \\ \gseq{\frames}{\vec{\Psi}}{\bnd{\Phi}{e_1}}}
  {\gseq{\frames}{\vec{\Psi}}{\bnd{\Phi}{e_2}}}
\and
\inferrule*[lab=pure-intro]
  {P \text{ holds in Lean}}
  {\gseq{\vec{\Gamma}}{\vec{\Psi}}{\gpure{P}}}
\end{mathpar}
\end{figrules}
\caption{Provability rules, fragment: the rules adapted to the frame structure of contexts, the rules for the later modality, and the rules that connect the logic to Lean and to the equational judgment.
The remaining rules are those of natural deduction for higher-order logic.
$\later\Phi$ abbreviates $\elift{(\vdelay{\Phi})}$ and $\gpure{P}$ abbreviates $\syn{pure}\;\delta(P)$.
Well-typedness side conditions are omitted.}
\label{fig:proves}
\end{figure*}

The rules of the logic are those of natural deduction for higher-order logic, adapted to the frame structure of contexts at two points.
First, the rule \textsc{asm} uses a hypothesis from frame $n$ as $\act{\Phi}{\wkf{n}}$.
That is, hypotheses cross frames by weakening, and what was provable $n$ ticks ago remains provable now.
As with $\eadv{n}{}$, building the weakening into the rule avoids explicit structural rules for the stack~\citep{DBLP:journals/jfp/HuJP23}.
Second, the rules that introduce a variable, namely \textsc{forall-intro} and \textsc{exists-elim}, extend the innermost variable frame and pad the hypothesis stack with $\wkv$, which keeps the two stacks aligned.

\paragraph{Later}
We write $\later\Phi$ for $\elift{(\vdelay{\Phi})}$.
The rule \textsc{lift-intro} moves into a fresh frame: in order to prove $\later\Phi$ now, we must prove $\Phi$ one tick later, with the whole context still available but shifted by one frame.
\textsc{later-mono} is the monotonicity rule: from $\later P$, together with a proof of $Q$ in the next frame that may use $P$ as its only new hypothesis, one concludes $\later Q$.
The modality commutes with conjunction and disjunction (\textsc{later-and}, \textsc{later-or}) and distributes over universal quantification in the direction $\later\forall x.\,\Phi \vdash \forall x.\,\later\Phi$. 
The rule \textsc{delay-eq} states that to prove equality of delayed terms it suffices to prove equality of values produced later.

\paragraph{L\"ob induction}
Guarded recursion at the level of proofs is the rule \textsc{loeb-ind}: in order to prove $\Phi$, one may hypothesize $\later(\act{\Phi}{\wkf{1}})$, that is, the statement itself one tick later.

\paragraph{Lean}
Two rules connect the logic to Lean. From a Lean proof of $P$, the
rule \textsc{pure-intro} proves $\gpure{P}$, which abbreviates
$\syn{pure}\;\delta(P)$. The rule \textsc{forall-intro-points} proves
$\forall_{\Delta\,T}\,\Phi$ from a Lean proof of
$\bnd{\Phi}{\delta(a)}$ for every $a \col T$; it is sound because
$\Delta\,T$ is interpreted as a constant presheaf.

\paragraph{Equality} 
Propositional equality is backed by a separate judgment $\typed{\vec{\Gamma}}{e_1 \equiv e_2}{\type}$ of judgmental equality.
It enters the logic through \textsc{eq-def}, which is the reflexivity rule for propositional equality. Propositional equality is 
eliminated by \textsc{eq-elim}, \ie{}, by Leibniz's rule through a substitution $\bnd{\Phi}{e}$.
\Cref{fig:eq} shows the guarded fragment of the rules.
Altogether they make the judgment an equivalence relation, provide $\beta$-laws for each type former and $\eta$-laws for functions, products and the later modality, and include a congruence rule for every term former, including $\syn{delay}$, $\syn{adv}$ and $\syn{fix}$.
In the $\beta$ rule for $\tlater{}$, the weakening $\wkdelay{n-1}{|\Gamma_1|}$ carries the term $e$ to the expanded context, 
and the side condition of the rule mirrors the typing of $\syn{adv}$.
The congruence rules, including under ticks, allow our meta-level normalizer to rewrite anywhere in a term (\Cref{sec:canonical}).
Judgmental equality is sound with respect to the model, and derivability of an equation implies that both sides are well-typed.

Two rules go beyond the $\beta$, $\eta$ and congruence rules: \textsc{unfold}, which unfolds term-level fixed points, 
and \textsc{ax}, which equates two terms whose interpretations in the model coincide.
We use the latter rule in three places: for the equations between $\syn{MK}$ and $\syn{PROJ}$ generated by \lstinline|gtype|; 
for function extensionality (\lstinline|gfunext|), which we prove in the model; and for equating two quoted programs with equal denotations, which is how we reason about Mathlib morphisms inside \grean (\Cref{sec:extraction}, \Cref{sec:delay}).
Everything else is derived within the logic.

\begin{figure*}
\begin{figrules}
\begin{mathpar}
\inferrule*[lab=beta-delay]
  {n > 0 \\ m = |\Gamma_1| \\ \typed{\cdot ; \vec{\Gamma}}{e}{\type}}
  {\typed{\Gamma_1;\dots;\Gamma_n;\vec{\Gamma}}{\eadv{n}{(\vdelay{e})} \equiv \act{e}{\wkdelay{n-1}{m}}}{\type}}
\and
\inferrule*[lab=eta-delay]
  {\typed{\vec{\Gamma}}{e}{\tlater{\type}}}
  {\typed{\vec{\Gamma}}{e \equiv \vdelay{(\eadv{1}{e})}}{\tlater{\type}}}
\and
\inferrule*[lab=unfold]
  {\typed{(\tlater{\type} , \Gamma) ; \vec{\Gamma}}{e}{\type}}
  {\typed{\Gamma ; \vec{\Gamma}}{\syn{fix}\;e \equiv \bnd{e}{\vdelay{\act{(\syn{fix}\;e)}{\wkf{1}}}}}{\type}}
\and
\inferrule*[lab=cong-delay]
  {|\vec{\Gamma}| > 0 \\ \typed{\cdot ; \vec{\Gamma}}{e \equiv e'}{\type}}
  {\typed{\vec{\Gamma}}{\vdelay{e} \equiv \vdelay{e'}}{\tlater{\type}}}
\and
\inferrule*[lab=cong-adv]
  {n > 0 \\ \typed{\vec{\Gamma}}{e \equiv e'}{\tlater{\type}}}
  {\typed{\Gamma_1;\dots;\Gamma_n;\vec{\Gamma}}{\eadv{n}{e} \equiv \eadv{n}{e'}}{\type}}
\and
\inferrule*[lab=cong-fix]
  {\typed{(\tlater{\type} , \Gamma) ; \vec{\Gamma}}{e \equiv e'}{\type}}
  {\typed{\Gamma ; \vec{\Gamma}}{\syn{fix}\;e \equiv \syn{fix}\;e'}{\type}}
\and
\inferrule*[lab=beta-embed-apply]
  {|\vec{\Gamma}| > 0}
  {\typed{\vec{\Gamma}}{\delta(f) \odot \delta(x) \equiv \delta(f\;x)}{\Delta\,S}}
\and
\inferrule*[lab=ax]
  {\typed{\vec{\Gamma}}{e}{\type} \\ \typed{\vec{\Gamma}}{e'}{\type} \\ \sem{e} = \sem{e'}}
  {\typed{\vec{\Gamma}}{e \equiv e'}{\type}}
\end{mathpar}
\end{figrules}
\caption{The equational judgment, written $\typed{\vec{\Gamma}}{e_1 \equiv e_2}{\type}$; fragment: the rules for the modality and for $\syn{fix}$, the congruence rules under ticks, and the two rules that import Lean values and equalities of the model.
Not shown are the equivalence rules, the $\beta$- and $\eta$-rules for functions and products, the $\beta$-rules for sums, and the pointwise congruence rules.
In \textsc{beta-embed-apply}, $f \col T \to S$ and $x \col T$ are Lean values. 
In \textsc{ax}, the third premise is equality of the two interpretations in the model.
}
\label{fig:eq}
\end{figure*}

\subsection{Soundness and extraction}
\label{sec:soundness}

Sequents are interpreted in the internal logic of $\topos$.
A proposition over $\vec{\Gamma}$ denotes a morphism $\sem{\vec{\Gamma}} \longto \Omega$, and a hypothesis stack denotes a list of such morphisms. Since these are defined using the interpretation of terms, they are partial functions, but well-formedness
of the judgment $\vec\Gamma \vdash \vec \Psi$ implies that $\sem{\vec{\Psi}}$ is well-defined. 
Internal entailment $P \ient Q$ holds when, at every stage $n$ and for every element $\gamma$ of $\sem{\vec{\Gamma}}$ at that stage, the sieve $P\,\gamma$ is contained in the sieve $Q\,\gamma$.

\begin{theorem}[Soundness]
If $\gseq{\vec{\Gamma}}{\vec{\Psi}}{\Phi}$ and $\vec{\Gamma} \vdash \vec{\Psi}$ then $\sem{\vec{\Psi}} \ient \sem{\Phi}$.
\end{theorem}

The proof is by induction on derivations. The case of \textsc{eq-def} uses soundness of the judgmental equality.
As a corollary we obtain consistency: $\bot$ is not derivable in the empty context.

\phantomsection
\label{sec:extraction}

Soundness is used to export \grean theorems into Lean. 
When exporting closed theorems, we simply use the fact that $\vec{\Gamma} \vdash \vec{\Psi}$ holds vacuously for empty $\vec{\Gamma}$ and $\vec{\Psi}$. 
A closed proposition $P$ is \emph{semantically valid} when
\[
  \Valid\;P \;\eqdef\; \sem{P}\ \text{is defined and}\ \top \ient \sem{P},
\]
with some rules shown below:
\begin{mathpar}
\inferrule*[lab=of-goal]
  {\gseq{\cdot}{\cdot}{P}}
  {\Valid\;P}
\and
\inferrule*[lab=pure]
  {\Valid\;(\gpure{\varphi})}
  {\varphi}
\and
\inferrule*[lab=later]
  {\Valid\;(\later P)}
  {\Valid\;P}
\end{mathpar}

The rule \textsc{of-goal} is soundness at the empty context, \textsc{later} allows one to peel $\later$ off a valid proposition, and \textsc{pure} is the exit: a valid $\syn{pure}\;\delta(\varphi)$ yields the Lean proposition $\varphi$ itself.
The rule \textsc{later} may look suspicious, since $\later\Phi \vdash \Phi$ is not provable in the logic.
It is sound nevertheless, because validity quantifies over all stages: $\later P$ at stage $n+1$ is $P$ at stage $n$, and therefore a valid $\later P$ gives a valid $P$.
The ticks accumulated in the proof are discarded on the way out, and an extracted theorem does not mention the deep embedding of the syntax or the provability judgment.
In \Cref{sec:adequacy} we use these rules to obtain an adequacy theorem stated in Lean.

 \subsection{The proof mode}
\label{sec:proofmode}

Working directly with a deep embedding requires explicit manipulations of syntax.
For instance, introducing a universally quantified variable requires weakening the hypotheses to account for the new variable.
\grean's proof mode provides a set of tactics that take care of these operations behind the curtains, allowing users to work with named variables and hypotheses, as in \Cref{fig:teaser-proof}.

\paragraph{Simplification}
\phantomsection
\label{sec:canonical}

The simplifier contracts $\beta$-redexes using \grean's equational theory.
For a term $e$ of type $\type$, it returns a term $e'$ together with a derivation of $\typed{\vec{\Gamma}}{e \equiv e'}{\type}$.
The reductions include function application, product projections, case expressions, and cancellation of $\syn{adv}$ against $\syn{delay}$; fixpoints are not unfolded by the simplifier, so that repeated simplification terminates.
The tactic \tac{gsimpl} repeats simplification until no reduction applies outside fixpoint bodies.
Unfolding a fixpoint is implemented by a separate tactic: \tac{gfix} applies \textsc{unfold} of \Cref{fig:eq} to exposed fixpoints, that is, those not under $\syn{delay}$, $\syn{adv}$ or another fixpoint, and then simplifies, so that the $\syn{adv}_{1}$ in the body cancels the $\syn{delay}$ around the unfolded definitions.

These reductions, and rules that extend the context, introduce weakenings and substitutions.
We push them to variables and quotations using the syntactic laws of \Cref{sec:weakening}.
We also use this mechanism with the other tactics that produce weakenings and substitutions: those that extend the context or enter a frame and weaken the hypotheses (\eg{}, \tac{gintro}), and those that instantiate bound variables by substitution (\eg{}, \tac{gapply}).
Quotations are handled as described in \Cref{sec:quotation}.

\paragraph{Rewriting}
\phantomsection
\label{sec:rewriting}
One of the most used tactics in our framework, \tac{grewrite h}, replaces an occurrence of one side of an equation \tac{h} in the goal by the other side, like Lean's \tac{rw}.
The step is justified by \textsc{eq-elim}, which needs a \emph{motive}: the goal with the occurrence abstracted into a fresh variable.
In \grean, constructing the motive is slightly different from non-modal languages, because the occurrence may be under $\syn{delay}$ and $\syn{adv}$, and the fresh variable must be referred to across frames.

Consider \tac{grewrite Htlf} in \Cref{fig:teaser-proof}.
If we write $a$ for the recursive call, \tac{Htlf} states $\vdelay{a} = \vdelay{(\eadv{1}{v})}$.
The tactic abstracts the occurrence of $\vdelay{a}$ in the goal, producing a \emph{motive}: a proposition in the context extended by a fresh variable.
In named notation, the resulting motive is
\[
  M(z) := \quo{\syn{Delay.MK}\,A}\,(\syn{inr}\;z) = d
\]
where $z$ names the fresh variable $\syn{var}\,(0,0)$.
The original goal contained $\bnd{M}{\vdelay{a}}$ and the new goal has $\bnd{M}{\vdelay{(\eadv{1}{v})}}$.
The tactic reduces the former to the latter, so it applies \textsc{eq-elim} to the new goal with the equation $\vdelay{(\eadv{1}{v})} = \vdelay{a}$.

Here the occurrence is at the top level.
Had we rewritten $a$ itself, under the $\syn{delay}$, the motive would be 
\[
  \quo{\syn{Delay.MK}\,A}\,(\syn{inr}\;(\vdelay{\syn{var}\,(1,0)})) = d,
\] 
with the fresh variable one frame back, because it is located under $\syn{delay}$.
The tactic implementation traverses the term to find the occurrence of a term to rewrite, while keeping a tick depth $d$, initially $0$, and a renaming $\ren$, initially $\wkv$, which is the weakening by the fresh variable as seen from the current position.
Under $\syn{delay}$, $d$ increases by one and $\ren$ becomes $\rliftf{\ren}$.
Under $\eadv{n}{}$ with $n\le d$, the offset becomes $\offset{\ren}{n}$, and the traversal continues at depth $d-n$ with renaming $\cut{\ren}{n}$.
When $n>d$, the operand belongs to a frame preceding the fresh variable, so the traversal does not enter it.
Next, a matching subterm is replaced by $\syn{var}\,(d,0)$.
Non-binding constructors are traversed structurally.
Subterms containing no replacement are renamed by $\ren$.
Finally, the tactic typechecks the motive and proves that substituting the two sides recovers the original and rewritten propositions.
When arguments of an equation are omitted, a structural search tries the target and then its subterms, taking the first match that determines all remaining arguments.
Quotations are compared by their $\SYNT$ values.

\paragraph{Logical tactics.}
For a goal $\later Q$ and selected hypotheses $\later P_i$, \tac{gmono} opens a fresh frame with goal $Q$ and hypotheses $P_i$ under the supplied names.
Several hypotheses are combined under $\later$ into a conjunction, which is split after applying \textsc{later-mono}, as described in \Cref{sec:rules}.
For a goal $\vdelay{a}=\vdelay{b}$, the tactic first applies \textsc{delay-eq}.
This is how \tac{gmono IH as G} makes the induction hypothesis available for the recursive call in \Cref{fig:teaser-proof}.
The tactic \tac{gnext} enters the next frame without unboxing a hypothesis.
Introduction, case analysis and intermediate assertions use \tac{gintro}, \tac{gcases} and \tac{gassert}; the case patterns follow Lean's \tac{rcases}, with branch equations for sum-typed terms.
The tactic \tac{gpoints a} for a goal $\forall_{\Delta\,T}\,\Phi$ introduces a Lean variable $a\col T$ and leaves $\bnd{\Phi}{\delta(a)}$ as the \grean goal, with the \grean contexts unchanged.
The tactic \tac{gembed} applies \textsc{pure-intro}, reducing a \grean goal $\syn{pure}\;\delta(P)$ to the Lean goal $P$.

\paragraph{Implementation.}
\label{sec:implementation}

\grean's provability judgment is implemented as an inductive family
\lstinline|PROVES : CTX → PCTX → EXPR → Prop|,
which in turn is used by a wrapper with additional indices that record variable and hypothesis names for elaboration and display.
The tactics are Lean metaprograms using Qq~\citep{qq} to construct typed Lean expressions.
They share operations for looking up hypotheses, weakening them into the current context, and constructing the corresponding derivations.
Typing side conditions are discharged by the typechecker and arithmetic conditions by decision procedures.
Lean's kernel checks the assembled proof term. \section{Case studies}
\label{sec:eval}

In this section, we apply \grean to two case studies: We first return
to the delay monad of \Cref{sec:overview} to prove its monad laws and
to characterize it as a free delay algebra. We then interpret a
$\lambda$-calculus with fixpoints in a guarded recursive domain and
prove soundness and adequacy. None of these results are new~\cite{DBLP:journals/entcs/PaviottiMB15}, but they
illustrate the expressive power of \grean.

\subsection{The guarded delay monad}
\label{sec:delay}

We extend the example of \Cref{sec:overview} with definitions for the left inclusion
$\syn{ret}\col A \to \Delay\,A$, the right inclusion $\syn{step} \col
\tlater{(\Delay\,A)}\to \Delay\,A$, and the monadic bind operation $\syn{bind}\col  (A \to \Delay\,
B) \to \Delay\, A \to \Delay\,B$. We prove the functor equations for
$\syn{map}$ and the monad equation for $\syn{bind}$ and $\syn{ret}$ by
unfolding their guarded definitions. The left unit law is simply the
definition of $\syn{bind}$ on $\syn{ret}$, while right unit,
associativity and preservation of composition by $\syn{map}$ use L\"ob
induction and case analysis similarly to the proof in
\Cref{fig:teaser-proof}.

Interpreting these operations on $\Delay$ gives us a monad on $\topos$.
Showing that $\Delay$ satisfies the monadic laws uses both connections between \grean and the model: $\syn{ax}$ and soundness, in a round trip.
Consider, for example, naturality of the unit.
A morphism from the topos of trees, $f \from X \longto Y$, is identified with its transpose $\One \longto Y^X = \sem{\tarr{\syn{ax}\;X}{\syn{ax}\;Y}}$, as the closed term $\overline{f} \eqdef \syn{ax}_{\tarr{\syn{ax}\;X}{\syn{ax}\;Y}}\;f$.
We must show $\sem{\syn{map}\;\overline{f}} \circ \sem{\syn{ret}_X} = \sem{\syn{ret}_Y} \circ f$, where the subscript records the type argument $\syn{ax}\;X$ or $\syn{ax}\;Y$ of $\syn{ret}$.
At this point, we treat $\overline{f}$ as an opaque constant in \grean.
We then prove the closed equation $\syn{ret}_Y \circ \overline{f} = \syn{map}\;\overline{f} \circ \syn{ret}_X$ between functions $\tarr{\syn{ax}\;X}{\Delay\,(\syn{ax}\;Y)}$ in \grean : by function extensionality it suffices to show $\syn{ret}_Y\;(\overline{f}\;x) = \syn{map}\;\overline{f}\;(\syn{ret}_X\;x)$, which is the equation of $\syn{map}$ on $\syn{ret}$ obtained by unfolding the fixpoint.
Soundness exports the equation as an equality of denotations using two facts: the denotation of a composite is the composite of the denotations, and $\sem{\overline{f}} = f$.
The other fields of the Mathlib \tac{CategoryTheory.Monad} on $\topos$ are obtained in the same way.

\subsection{Delay is the free delay algebra}
\label{sec:freealg}

A \emph{delay algebra} is an object $B$ with a map $s\col\tarr{\tlater{B}}{B}$.
For $f\col\tarr{A}{B}$, guarded recursion defines an extension $\syn{ext}\;s\;f\col\tarr{\Delay\,A}{B}$ satisfying
\[
  \begin{array}{l@{\;}l}
  \syn{ext}\;s\;f\;(\syn{ret}\;a) &= f\;a \\
  \syn{ext}\;s\;f\;(\syn{step}\;w) &= s\;(\vdelay{\syn{ext}\;s\;f\;(\eadv{1}{w})}).
  \end{array}
\]
These equations can be used to express $\syn{ext}\;s\;f\col\tarr{\Delay\,A}{B}$ as a fixed point.
For uniqueness, suppose $h$ is another map satisfying the above equations.
We prove $\forall d.\ h\;d=\syn{ext}\;s\;f\;d$ by L\"ob induction and case analysis on $\syn{PROJ}\;d$.
The return case follows from the first equation.
The step case reduces to equality of delayed recursive calls, where \tac{gmono} makes the induction hypothesis available.
Function extensionality then gives $h=\syn{ext}\;s\;f$. Freeness provides an alternative route to showing 
that $\Delay$ is a monad.

\subsection{A lambda calculus with fixpoints}
\label{sec:adequacy}

\newcommand{\lamfix}{$\lambda_{\syn{fix}}$\xspace} 

In this example, we define the model for a simply-typed call-by-name $\lambda$-calculus
with unit and fixpoints, and prove it computationally adequate. Since \garlene\ does not have 
inductive types, we represent syntax, typing judgements and the operational semantics
using Lean types. Writing \lamfix for this
object language, we define an inductive family $\Lam\,n$ of \lamfix terms in
the scope of $n$ variables, an inductive type of \lamfix types
$a\cceq\syn{unit}\mid\tarr{a}{b}$, and a typing judgment
$\syn{Typing}\;\Gamma\;e\;a$. We write $e\to e'$ for one reduction
step and $e\opsteps e'$ for its reflexive-transitive closure.
The grammar and operational semantics of \lamfix are shown below. 
$i\col\syn{Fin}\,n$ is a de Bruijn index, the bodies under binders are terms of type $\Lam\,(n+1)$ and $t[u]$ substitutes $u$ for the variable $0$ of $t$.
\[
  e \cceq \syn{unit} \mid \syn{var}\;i \mid \syn{app}\;e\;e' \mid \syn{lam}\;e \mid \syn{fix}\;e
\]
\[
  \begin{array}{@{}l@{\;}c@{\;}l@{\qquad}l@{}}
    \syn{app}\;(\syn{lam}\;t)\;u &\to& t[u] \\
    \syn{app}\;t\;u &\to& \syn{app}\;t'\;u & \text{if } t\to t' \\
    \syn{fix}\;t &\to& t[\syn{fix}\;t]
  \end{array}
\]

We interpret \lamfix in the guarded type $\PCF$ of \Cref{sec:rectypes}.
Its four summands represent numbers, an error, delayed computations,
and delayed functions. 
We write $\syn{num}\;n$, $\syn{error}$, $\syn{thunk}\;w$ and $\syn{lam}\;g$ for the four injections into $\PCF$.
An environment $\syn{Env}\,n$ contains $n$ elements of
$\PCF$. The interpretation
$\syn{interp}_n\col\Lam\,n\to\SYNT\,(\tarr{\syn{Env}\,n}{\PCF})$ is defined by recursion on the \lamfix syntax in Lean, using guarded recursion to interpret \lamfix fixpoints. 
The interpretation is given by the following clauses, omitting quotations, where $\rho\col\syn{Env}\,n$ is an environment and $\rho_i$ is its $i$ entry:
\[
  \begin{array}{@{}l@{\;}c@{\;}l@{}}
    \syn{interp}_n\,\syn{unit}\;\rho &=& \syn{num}\;0 \\
    \syn{interp}_n\,(\syn{var}\;i)\;\rho &=& \rho_i \\
    \syn{interp}_n\,(\syn{app}\;t\;u)\;\rho &=& \syn{apply}\;(\syn{interp}_n\,t\;\rho)\;(\syn{interp}_n\,u\;\rho) \\
    \syn{interp}_n\,(\syn{lam}\;t)\;\rho &=& \syn{lam}\,(\vdelay{(\syn{\lambda}v\synds \syn{interp}_{n+1}\,t\;\vpair{\rho}{v}))} \\
    \syn{interp}_n\,(\syn{fix}\;t)\;\rho &=& \syn{fix}\;y\synds \\
    \multicolumn{3}{@{}l@{}}{\qquad \syn{thunk}\,(\vdelay{(\syn{interp}_{n+1}\,t\;\vpair{\rho}{\eadv{1}{y}})}).}
  \end{array}
\]
Application $\syn{apply}\col\PCF\to\PCF\to\PCF$ is defined in \grean by guarded recursion.
It returns $\syn{error}$ on $\syn{error}$ and $\syn{num}\;n$, and otherwise satisfies
\[
  \begin{array}{@{}l@{\;}c@{\;}l@{}}
    \syn{apply}\;(\syn{lam}\;g)\;x &=& \syn{thunk}\,(\vdelay{(\eadv{1}{g}\;x)}) \\
    \syn{apply}\;(\syn{thunk}\;w)\;x &=& \syn{thunk}\,(\vdelay{(\syn{apply}\;(\eadv{1}{w})\;x)}).
  \end{array}
\]
The soundness property states
that each \lamfix reduction $e\to e'$ corresponds to one thunk
in the model:
\[
  \gseq{\cdot}{\cdot}{\forall\rho.\ \syn{interp}_n\,e\;\rho}
  =\syn{thunk}\,(\vdelay{(\syn{interp}_n\,e'\;\rho)}).
\]
The proof is by induction on the reduction derivation, using substitution lemmas for the interpretation. In other words,
using Lean induction over the step relation we build a 
\garlene\ proof for each possible step using the \garlene\ proof mode. 

\paragraph{Adequacy}
To prove adequacy of our denotational semantics, we relate closed \lamfix terms to the model.
Given a relation $R$ between \lamfix terms and elements of $\PCF$, we define a relation $\syn{Exp}\;R\;e\;p$ that requires a \lamfix reduct related to each value exposed by $p$.
Omitting quotations and \lamfix term embeddings, we define
\[
  \syn{Exp}\;R\;e\;p\eqdef
  \syn{wp}\;p\;(\vlam{v}{\exists w.\ \gpure{e\opsteps w}\wedge R\;w\;v}).
\]The weakest precondition predicate $\syn{wp}$ is
defined by guarded recursion. It applies its postcondition to numbers
and functions, is false on $\syn{error}$, and postpones the condition
by one tick at a thunk:
\[
  \syn{wp}\;(\syn{thunk}\;t)\;\Phi
  =\later\bigl(\syn{wp}\;(\eadv{1}{t})\;\Phi\bigr).
\]
We define a logical relation $\syn{Val}_a$ by recursion on \lamfix types and use it as $R$ in the $\syn{Exp}$ relation.
At $\syn{unit}$, it requires $w=\syn{unit}$ and $p=\syn{num}\;0$.
At an arrow type, $p$ must be $\syn{lam}\;g$; applications to related arguments must be related one tick later, when $g$ is available:
\[
  \begin{array}{@{}l@{}}
    \syn{Val}_{\tarr{a}{b}}\;w\;p\eqdef
    \exists g.\ p=\syn{lam}\;g\ \wedge \\
    \quad \forall y.\ \later\bigl(\forall q.\ \syn{Exp}\;\syn{Val}_a\;y\;q \\
    \qquad {}\to\syn{Exp}\;\syn{Val}_b\;(\syn{app}\;w\;y)\;(\eadv{1}{g}\;q)\bigr).
  \end{array}
\]
The relation $\syn{SubstOk}\;\Gamma\;\sigma\;\rho$ between a closing substitution $\sigma\col\syn{Fin}\,n\to\Lam\,0$ and an environment $\rho\col\syn{Env}\,n$ is defined by recursion on $n$; for each variable $i$ it relates $\sigma\,i$ to the $i$-th entry of $\rho$ by $\syn{Exp}\;\syn{Val}_{\Gamma\,i}$.
The fundamental lemma relates every well-typed term to its interpretation under related substitutions and environments. 
\begin{lemma}[Fundamental lemma]
  For every typing context $\Gamma$ of length $n$, term $e\col\Lam\,n$ and type $a$ with $\syn{Typing}\;\Gamma\;e\;a$, and every closing substitution $\sigma\col\syn{Fin}\,n\to\Lam\,0$,
  \[
    \begin{array}{@{}l@{}}
      \gseq{\cdot}{\cdot}{\forall\rho\col\syn{Env}\,n.\ \syn{SubstOk}\;\Gamma\;\sigma\;\rho} \\
      \qquad {}\to\syn{Exp}\;\syn{Val}_a\;(e[\sigma])\;(\syn{interp}_n\,e\;\rho).
    \end{array}
  \]
\end{lemma}
We prove it by induction on typing derivations in Lean, carrying out each case in the proof mode.
In the fixpoint case, the typing induction hypothesis applies to the body, while L\"ob induction assumes the property of the recursive term one tick later.
Unfolding exposes a thunk; under it, \tac{gmono} gives the relation needed for the recursive environment entry.

We write $\syn{denote}\;e$ for the closed interpretation of $e$ in $\PCF$, 
and $\syn{step}\;r$ for $\syn{thunk}\,(\vdelay{\quo{r}})$ where $r : \PCF$. 
The Lean predicate $\syn{Runs}\;k\;p\;q$ states that $p$ is $q$ delayed by $k$ times $\syn{step}$: 
\[
  \begin{array}{@{}l@{\;}c@{\;}l@{}}
    \syn{Runs}\;0\;p\;q &\eqdef& (\gseq{\cdot}{\cdot}{\quo{p}=\quo{q}}) \\ \syn{Runs}\,(k{+}1)\;p\;q &\eqdef& \exists r.\ \syn{EqG}\;p\;(\syn{step}\;r) \wedge\syn{Runs}\;k\;r\;q.
  \end{array}
\]
Adequacy then states that if the denotation of a term terminates, also the term itself terminates.
\begin{theorem}[Adequacy]
  For every closed term $e\col\Lam\,0$ with $\syn{Typing}\;\Gamma\;e\;\syn{unit}$ and every $k\in\mathbb{N}$,
  \[
    \syn{Runs}\;k\;(\syn{denote}\;e)\;(\syn{num}\;0)
    \quad\Longrightarrow\quad e\opsteps\syn{unit}.
  \]
\end{theorem}

The proof uses the validity rules described in \Cref{sec:extraction}.
Soundness of \grean makes the closed instance of the fundamental lemma valid, and by induction on $k$ we transport validity of the fundamental lemma across the equations given by $\syn{Runs}$.
At $\syn{num}\;0$, the logical relation implies the valid proposition $\gpure{e\opsteps\syn{unit}}$, which we can extract to Lean and finish the proof.

The adequacy allows us to relate proofs about denotational semantics of \lamfix back to the operational semantics of \lamfix stated as usual inductive relation in Lean.

 \section{Conclusion and future work}
\label{sec:conclusion}

\garlene aims to serve as a practical tool for programming and
reasoning with guarded recursion. Learning from previous work on
embedding multimode type theories in
Agda~\citep{DBLP:journals/corr/abs-2207-00843,DBLP:journals/pacmpl/CeulemansND25},
we focussed on providing intuitive syntax and good performance that
scales to larger applications. Our experience -- documented in the two
case studies above -- suggests that \garlene is capable of mechanising
larger formal developments using guarded recursion. We hope that
\garlene will find use for mechanising existing and future work that
uses guarded type theory as metalanguage.

We envision that future work can build on the foundation laid by the
current implementation of \garlene to meaningfully extend it while
maintaining or improving upon its favourable ergonomics. For example,
\garlene internally uses a Hofmann--Streicher universe to construct
guarded recursive types declared via \lstinline|gtype|. By exposing
this universe as a type in \garlene, we could instead express guarded
recursive types directly as guarded fixed
points~\citep{DBLP:conf/lics/BirkedalM13}. 

We also plan to explore the use of \garlene as a language for programming
and reasoning about coinductive types. One way to do this could be to 
expand \garlene from the single clocked version to multiple 
clocks~\citep{atkey13ProductiveCoprogrammingGuarded}. Another would
be to use the denotational model, exploring the fact that many coinductive 
types can be expressed as sets of 
global elements of guarded recursive types in the topos of trees. 
For these applications, it will be interesting to extend the model of guarded recursion, 
by increasing the indexing to ordinals larger than $\omega$. This should
allow for programming and reasoning also about coinductive types 
whose definitions involve constructions such as finite or countable powersets
or finite distributions~\citep{mogelberg:beyondOmega}.
Such a model could also be used to reason about liveness properties
as in Transfinite Iris~\cite{TransfiniteIris}.

Finally, apart from presenting \garlene, this paper also describes key
implementation techniques that enabled the performance and usability
characteristics of \garlene. We hope that these insights can be
valuable for future work on embedding domain-specific calculi in
general-purpose proof assistants.

\paragraph{Use of LLMs}
Large language models were used to fill out some of the proofs in the Lean mechanization, primarily the case-heavy renaming and substitutions proofs used for simplification, and for debugging tactics. 
All relevant theorem statements were written by hand, so we do not expect risk of the LLMs introducing errors.

\section*{Data Availability Statement}
The Lean mechanization accompanying this work is available on Zenodo~\citep{zenodo} and on GitHub at
\url{https://github.com/Kaptch/Garlene}.

\section*{Acknowledgments}
This work was supported by the Independent Research Fund Denmark grant number 2032-00134B.

\label{sec:bib}

\appendix
\onecolumn
\section{Complete rule sets}
\label{app:rules}
\captionsetup{belowskip=6pt}

\Cref{fig:typing,fig:proves,fig:eq} show fragments of the three rule sets of the DSL.
This appendix lists them in full: the typing rules of the object language in \Cref{fig:typing-full}, the provability rules of the logic in \Cref{fig:proves-full}, and the equational judgment in \Cref{fig:eq-full}.

\begin{figrules}
\begin{mathpar}
\inferrule*[lab=embed]
  {|\vec{\Gamma}| > 0}
  {\typed{\vec{\Gamma}}{\delta(a)}{\Delta\,T}}
\and
\inferrule*[lab=embed-apply]
  {\typed{\vec{\Gamma}}{f}{\Delta(T \to S)} \\ \typed{\vec{\Gamma}}{x}{\Delta\,T}}
  {\typed{\vec{\Gamma}}{f \odot x}{\Delta\,S}}
\and
\inferrule*[lab=pure]
  {\typed{\vec{\Gamma}}{e}{\Delta\,\Prop}}
  {\typed{\vec{\Gamma}}{\syn{pure}\;e}{\Omega}}
\and
\inferrule*[lab=ax]
  {|\vec{\Gamma}| > 0}
  {\typed{\vec{\Gamma}}{\syn{ax}_{\type}\;f}{\type}}
\and
\inferrule*[lab=var]
  {\Gamma_p = \tau_0,\dots,\tau_q,\Gamma}
  {\typed{\Gamma_0;\dots;\Gamma_p;\vec{\Gamma}}{\syn{var}\;(p,q)}{\tau_q}}
\and
\inferrule*[lab=app]
  {\typed{\vec{\Gamma}}{e_1}{\tarr{\type}{\sigma}} \\ \typed{\vec{\Gamma}}{e_2}{\type}}
  {\typed{\vec{\Gamma}}{\syn{app}\;e_1\;e_2}{\sigma}}
\and
\inferrule*[lab=lam]
  {\typed{(\type , \Gamma) ; \vec{\Gamma}}{e}{\sigma}}
  {\typed{\Gamma ; \vec{\Gamma}}{\syn{lam}_{\type}\;e}{\tarr{\type}{\sigma}}}
\and
\inferrule*[lab=delay]
  {|\vec{\Gamma}| > 0 \\ \typed{\cdot ; \vec{\Gamma}}{e}{\type}}
  {\typed{\vec{\Gamma}}{\vdelay{e}}{\tlater{\type}}}
\and
\inferrule*[lab=adv]
  {n > 0 \\ \typed{\vec{\Gamma}}{e}{\tlater{\type}}}
  {\typed{\Gamma_1;\dots;\Gamma_n;\vec{\Gamma}}{\eadv{n}{e}}{\type}}
\and
\inferrule*[lab=fix]
  {\typed{(\tlater{\type} , \Gamma) ; \vec{\Gamma}}{e}{\type}}
  {\typed{\Gamma ; \vec{\Gamma}}{\syn{fix}_{\tlater{\type}}\;e}{\type}}
\and
\inferrule*[lab=pair]
  {\typed{\vec{\Gamma}}{e_1}{\type} \\ \typed{\vec{\Gamma}}{e_2}{\sigma}}
  {\typed{\vec{\Gamma}}{\vpair{e_1}{e_2}}{\tprod{\type}{\sigma}}}
\and
\inferrule*[lab=proj]
  {\typed{\vec{\Gamma}}{e}{\tprod{\type}{\sigma}}}
  {\typed{\vec{\Gamma}}{\syn{proj}\;e\;\syn{L}}{\type}}
\and
\inferrule*[lab=inl]
  {\typed{\vec{\Gamma}}{e}{\type}}
  {\typed{\vec{\Gamma}}{\syn{inl}\;e}{\type \mathbin{\syn{\oplus}} \sigma}}
\and
\inferrule*[lab=case]
  {\typed{\vec{\Gamma}}{e}{\type \mathbin{\syn{\oplus}} \sigma} \\ \typed{\vec{\Gamma}}{f}{\tarr{\type}{\rho}} \\ \typed{\vec{\Gamma}}{g}{\tarr{\sigma}{\rho}}}
  {\typed{\vec{\Gamma}}{\syn{case}\;e\;f\;g}{\rho}}
\and
\inferrule*[lab=conn]
  {\typed{\vec{\Gamma}}{\Phi_1}{\Omega} \\ \typed{\vec{\Gamma}}{\Phi_2}{\Omega}}
  {\typed{\vec{\Gamma}}{\Phi_1 \bullet \Phi_2}{\Omega}}
\and
\inferrule*[lab=quant]
  {\typed{(\type , \Gamma) ; \vec{\Gamma}}{\Phi}{\Omega}}
  {\typed{\Gamma ; \vec{\Gamma}}{\mathrm{Q}_{\type}\,\Phi}{\Omega}}
\and
\inferrule*[lab=lift]
  {\typed{\vec{\Gamma}}{e}{\tlater{\Omega}}}
  {\typed{\vec{\Gamma}}{\elift{e}}{\Omega}}
\and
\inferrule*[lab=eq]
  {\typed{\vec{\Gamma}}{e_1}{\type} \\ \typed{\vec{\Gamma}}{e_2}{\type}}
  {\typed{\vec{\Gamma}}{e_1 =_{\type} e_2}{\Omega}}
\and
\inferrule*[lab=true/false]
  {|\vec{\Gamma}| > 0}
  {\typed{\vec{\Gamma}}{c}{\Omega}}
\end{mathpar}
\captionof{figure}{Typing rules, complete.
A context $\vec{\Gamma}$ is a stack of frames $\Gamma$ (\Cref{sec:ticks}).
\textsc{proj} is shown for $\syn{L}$; the $\syn{R}$ projection returns $\sigma$, and \textsc{inr} is dual to \textsc{inl}.
In \textsc{conn}, $\bullet$ ranges over $\wedge$, $\vee$, $\to$; in \textsc{quant}, $\mathrm{Q}$ ranges over $\forall$, $\exists$; in \textsc{true/false}, $c$ ranges over $\top$, $\bot$.
In \textsc{ax}, $f \from \One \longto \sem{\type}$ is a global element of the model.
Well-typedness side conditions carried by the mechanization are omitted except where they matter. }
\label{fig:typing-full}
\end{figrules}
\bigskip

\begin{figrules}
\begin{mathpar}
\inferrule*[lab=asm]
  {\Phi \in \Psi_n}
  {\gseq{\vec{\Gamma}}{\Psi_0;\dots;\Psi_n;\vec{\Psi}}{\act{\Phi}{\wkf{n}}}}
\and
\inferrule*[lab=true-intro]
  { }
  {\gseq{\vec{\Gamma}}{\vec{\Psi}}{\top}}
\and
\inferrule*[lab=false-elim]
  {\gseq{\vec{\Gamma}}{\vec{\Psi}}{\bot}}
  {\gseq{\vec{\Gamma}}{\vec{\Psi}}{\Phi}}
\and
\inferrule*[lab=and-intro]
  {\gseq{\vec{\Gamma}}{\vec{\Psi}}{\Phi_1} \\ \gseq{\vec{\Gamma}}{\vec{\Psi}}{\Phi_2}}
  {\gseq{\vec{\Gamma}}{\vec{\Psi}}{\Phi_1 \wedge \Phi_2}}
\and
\inferrule*[lab=and-elim]
  {\gseq{\vec{\Gamma}}{\vec{\Psi}}{\Phi_1 \wedge \Phi_2}}
  {\gseq{\vec{\Gamma}}{\vec{\Psi}}{\Phi_i}}
\and
\inferrule*[lab=or-intro]
  {\gseq{\vec{\Gamma}}{\vec{\Psi}}{\Phi_i}}
  {\gseq{\vec{\Gamma}}{\vec{\Psi}}{\Phi_1 \vee \Phi_2}}
\and
\inferrule*[lab=or-elim]
  {\gseq{\vec{\Gamma}}{(P , \Psi) ; \vec{\Psi}}{\Phi} \\ \gseq{\vec{\Gamma}}{(Q , \Psi) ; \vec{\Psi}}{\Phi} \\ \gseq{\vec{\Gamma}}{\Psi ; \vec{\Psi}}{P \vee Q}}
  {\gseq{\vec{\Gamma}}{\Psi ; \vec{\Psi}}{\Phi}}
\and
\inferrule*[lab=impl-intro]
  {\gseq{\vec{\Gamma}}{(\Phi_1 , \Psi) ; \vec{\Psi}}{\Phi_2}}
  {\gseq{\vec{\Gamma}}{\Psi ; \vec{\Psi}}{\Phi_1 \to \Phi_2}}
\and
\inferrule*[lab=impl-elim]
  {\gseq{\vec{\Gamma}}{\vec{\Psi}}{\Phi_1 \to \Phi_2} \\ \gseq{\vec{\Gamma}}{\vec{\Psi}}{\Phi_1}}
  {\gseq{\vec{\Gamma}}{\vec{\Psi}}{\Phi_2}}
\\
\inferrule*[lab=forall-intro]
  {\gseq{(\type , \Gamma) ; \frames}{\act{\vec{\Psi}}{\wkv}}{\Phi}}
  {\gseq{\Gamma ; \frames}{\vec{\Psi}}{\forall_{\type}\,\Phi}}
\and
\inferrule*[lab=forall-intro-points]
  {\forall a \col T.\, \gseq{\vec{\Gamma}}{\vec{\Psi}}{\bnd{\Phi}{\delta(a)}}}
  {\gseq{\vec{\Gamma}}{\vec{\Psi}}{\forall_{\Delta\,T}\,\Phi}}
\and
\inferrule*[lab=forall-elim]
  {\gseq{\Gamma ; \frames}{\vec{\Psi}}{\forall_{\type}\,\Phi} \\ \typed{\Gamma ; \frames}{e}{\type}}
  {\gseq{\Gamma ; \frames}{\vec{\Psi}}{\bnd{\Phi}{e}}}
\and
\inferrule*[lab=exists-intro]
  {\typed{\Gamma ; \frames}{e}{\type} \\ \gseq{\Gamma ; \frames}{\vec{\Psi}}{\bnd{\Phi}{e}}}
  {\gseq{\Gamma ; \frames}{\vec{\Psi}}{\exists_{\type}\,\Phi}}
\and
\inferrule*[lab=exists-elim]
  {\gseq{\Gamma ; \frames}{\Psi ; \vec{\Psi}}{\exists_{\type}\,\Phi} \\ \gseq{(\type , \Gamma) ; \frames}{(\Phi , \act{\Psi}{\wkv}) ; \vec{\Psi}}{\act{Q}{\wkv}}}
  {\gseq{\Gamma ; \frames}{\Psi ; \vec{\Psi}}{Q}}
\and
\inferrule*[lab=sum-elim]
  {\typed{\Gamma ; \frames}{e}{\type \mathbin{\syn{\oplus}} \sigma} \\\\
   \gseq{(\type , \Gamma) ; \frames}{(\act{e}{\wkv} = \syn{inl}\;x , \act{\Psi}{\wkv}) ; \vec{\Psi}}{\act{\Phi}{\wkv}} \\\\
   \gseq{(\sigma , \Gamma) ; \frames}{(\act{e}{\wkv} = \syn{inr}\;x , \act{\Psi}{\wkv}) ; \vec{\Psi}}{\act{\Phi}{\wkv}}}
  {\gseq{\Gamma ; \frames}{\Psi ; \vec{\Psi}}{\Phi}}
\and
\inferrule*[lab=lift-intro]
  {\gseq{\cdot ; \vec{\Gamma}}{\cdot ; \vec{\Psi}}{\Phi}}
  {\gseq{\vec{\Gamma}}{\vec{\Psi}}{\later\Phi}}
\and
\inferrule*[lab=forall-intro-points]
  {\forall a \col T.\, \gseq{\vec{\Gamma}}{\vec{\Psi}}{\bnd{\Phi}{\delta(a)}}}
  {\gseq{\vec{\Gamma}}{\vec{\Psi}}{\forall_{\Delta\,T}\,\Phi}}
\and
\inferrule*[lab=later-mono]
  {\gseq{\vec{\Gamma}}{\vec{\Psi}}{\later P} \\ \gseq{\cdot ; \vec{\Gamma}}{P ; \vec{\Psi}}{Q}}
  {\gseq{\vec{\Gamma}}{\vec{\Psi}}{\later Q}}
\and
\inferrule*[lab=later-and]
  {\gseq{\vec{\Gamma}}{\vec{\Psi}}{\later P} \\ \gseq{\vec{\Gamma}}{\vec{\Psi}}{\later Q}}
  {\gseq{\vec{\Gamma}}{\vec{\Psi}}{\later(P \wedge Q)}}
\and
\inferrule*[lab=later-or]
  {\gseq{\vec{\Gamma}}{\vec{\Psi}}{\later(P \vee Q)}}
  {\gseq{\vec{\Gamma}}{\vec{\Psi}}{\later P \vee \later Q}}
\and
\inferrule*[lab=loeb-ind]
  {\gseq{\vec{\Gamma}}{(\later(\act{\Phi}{\wkf{1}}) , \Psi) ; \vec{\Psi}}{\Phi}}
  {\gseq{\vec{\Gamma}}{\Psi ; \vec{\Psi}}{\Phi}}
\and
\inferrule*[lab=delay-eq]
  {\gseq{\vec{\Gamma}}{\vec{\Psi}}{\later(e_1 =_{\type} e_2)}}
  {\gseq{\vec{\Gamma}}{\vec{\Psi}}{\vdelay{e_1} =_{\tlater{\type}} \vdelay{e_2}}}
\and
\inferrule*[lab=eq-def]
  {\typed{\vec{\Gamma}}{e_1 \equiv e_2}{\type}}
  {\gseq{\vec{\Gamma}}{\vec{\Psi}}{e_1 =_{\type} e_2}}
\and
\inferrule*[lab=eq-elim]
  {\gseq{\Gamma ; \frames}{\vec{\Psi}}{e_1 =_{\type} e_2} \\ \gseq{\Gamma ; \frames}{\vec{\Psi}}{\bnd{\Phi}{e_1}}}
  {\gseq{\Gamma ; \frames}{\vec{\Psi}}{\bnd{\Phi}{e_2}}}
\and
\inferrule*[lab=pure-intro]
  {P \text{ holds in Lean}}
  {\gseq{\vec{\Gamma}}{\vec{\Psi}}{\syn{pure}\;\delta(P)}}
\and
\inferrule*[lab=inl-inj]
  {\gseq{\vec{\Gamma}}{\vec{\Psi}}{\syn{inl}\;a = \syn{inl}\;a'}}
  {\gseq{\vec{\Gamma}}{\vec{\Psi}}{a = a'}}
\and
\inferrule*[lab=inl-inr-disj]
  {\gseq{\vec{\Gamma}}{\vec{\Psi}}{\syn{inl}\;a = \syn{inr}\;b}}
  {\gseq{\vec{\Gamma}}{\vec{\Psi}}{\bot}}
\end{mathpar}
\captionof{figure}{Provability rules, complete.
In \textsc{and-elim} and \textsc{or-intro}, $i \in \{1,2\}$; \textsc{inl-inj} has an \textsc{inr} dual.
$\later\Phi$ abbreviates $\elift{(\vdelay{\Phi})}$. Well-typedness side conditions are omitted except where they matter.}
\label{fig:proves-full}
\end{figrules}
\bigskip

\begin{figrules}
\begin{mathpar}
\inferrule*[lab=rfl]
  {\typed{\vec{\Gamma}}{e}{\type}}
  {\typed{\vec{\Gamma}}{e \equiv e}{\type}}
\and
\inferrule*[lab=sym]
  {\typed{\vec{\Gamma}}{e_1 \equiv e_2}{\type}}
  {\typed{\vec{\Gamma}}{e_2 \equiv e_1}{\type}}
\and
\inferrule*[lab=tran]
  {\typed{\vec{\Gamma}}{e_1 \equiv e_2}{\type} \\ \typed{\vec{\Gamma}}{e_2 \equiv e_3}{\type}}
  {\typed{\vec{\Gamma}}{e_1 \equiv e_3}{\type}}
\and
\inferrule*[lab=beta-embed-apply]
  {|\vec{\Gamma}| > 0}
  {\typed{\vec{\Gamma}}{\delta(f) \odot \delta(x) \equiv \delta(f\;x)}{\Delta\,S}}
\and
\inferrule*[lab=beta-lam]
  {\typed{\Gamma ; \frames}{e'}{\type} \\ \typed{(\type , \Gamma) ; \frames}{e}{\sigma}}
  {\typed{\Gamma ; \frames}{\syn{app}\;(\syn{lam}_{\type}\;e)\;e' \equiv \bnd{e}{e'}}{\sigma}}
\and
\inferrule*[lab=eta-lam]
  {\typed{\vec{\Gamma}}{e}{\tarr{\type}{\sigma}}}
  {\typed{\vec{\Gamma}}{e \equiv \syn{lam}_{\type}\;(\syn{app}\;\act{e}{\wkv}\;(\syn{var}\;(0,0)))}{\tarr{\type}{\sigma}}}
\and
\inferrule*[lab=beta-delay]
  {n > 0 \\ m = |\Gamma_1| \\ \typed{\cdot ; \vec{\Gamma}}{e}{\type}}
  {\typed{\Gamma_1;\dots;\Gamma_n;\vec{\Gamma}}{\eadv{n}{(\vdelay{e})} \equiv \act{e}{\wkdelay{n-1}{m}}}{\type}}
\and
\inferrule*[lab=eta-delay]
  {\typed{\vec{\Gamma}}{e}{\tlater{\type}}}
  {\typed{\vec{\Gamma}}{e \equiv \vdelay{(\eadv{1}{e})}}{\tlater{\type}}}
\and
\inferrule*[lab=unfold]
  {\typed{(\tlater{\type} , \Gamma) ; \vec{\Gamma}}{e}{\type}}
  {\typed{\Gamma ; \vec{\Gamma}}{\syn{fix}\;e \equiv \bnd{e}{\vdelay{\act{(\syn{fix}\;e)}{\wkf{1}}}}}{\type}}
\and
\inferrule*[lab=beta-prod]
  {\typed{\vec{\Gamma}}{e}{\type} \\ \typed{\vec{\Gamma}}{e'}{\sigma}}
  {\typed{\vec{\Gamma}}{\syn{proj}\;\vpair{e}{e'}\;\syn{L} \equiv e}{\type}}
\and
\inferrule*[lab=eta-prod]
  {\typed{\vec{\Gamma}}{e}{\tprod{\type}{\sigma}}}
  {\typed{\vec{\Gamma}}{e \equiv \vpair{\syn{proj}\;e\;\syn{L}}{\syn{proj}\;e\;\syn{R}}}{\tprod{\type}{\sigma}}}
\and
\inferrule*[lab=beta-case]
  {\typed{\vec{\Gamma}}{a}{\type} \\ \typed{\vec{\Gamma}}{f}{\tarr{\type}{\rho}} \\ \typed{\vec{\Gamma}}{g}{\tarr{\sigma}{\rho}}}
  {\typed{\vec{\Gamma}}{\syn{case}\;(\syn{inl}\;a)\;f\;g \equiv \syn{app}\;f\;a}{\rho}}
\and
\inferrule*[lab=cong-lam]
  {\typed{(\type , \Gamma) ; \vec{\Gamma}}{e \equiv e'}{\sigma}}
  {\typed{\Gamma ; \vec{\Gamma}}{\syn{lam}_{\type}\;e \equiv \syn{lam}_{\type}\;e'}{\tarr{\type}{\sigma}}}
\and
\inferrule*[lab=cong-forall]
  {\typed{(\type , \Gamma) ; \vec{\Gamma}}{\Phi \equiv \Phi'}{\Omega}}
  {\typed{\Gamma ; \vec{\Gamma}}{\forall_{\type}\,\Phi \equiv \forall_{\type}\,\Phi'}{\Omega}}
\and
\inferrule*[lab=cong-fix]
  {\typed{(\tlater{\type} , \Gamma) ; \vec{\Gamma}}{e \equiv e'}{\type}}
  {\typed{\Gamma ; \vec{\Gamma}}{\syn{fix}\;e \equiv \syn{fix}\;e'}{\type}}
\and
\inferrule*[lab=cong-delay]
  {|\vec{\Gamma}| > 0 \\ \typed{\cdot ; \vec{\Gamma}}{e \equiv e'}{\type}}
  {\typed{\vec{\Gamma}}{\vdelay{e} \equiv \vdelay{e'}}{\tlater{\type}}}
\and
\inferrule*[lab=cong-adv]
  {n > 0 \\ \typed{\vec{\Gamma}}{e \equiv e'}{\tlater{\type}}}
  {\typed{\Gamma_1;\dots;\Gamma_n;\vec{\Gamma}}{\eadv{n}{e} \equiv \eadv{n}{e'}}{\type}}
\and
\inferrule*[lab=ax]
  {\typed{\vec{\Gamma}}{e}{\type} \\ \typed{\vec{\Gamma}}{e'}{\type} \\ \sem{e} = \sem{e'}}
  {\typed{\vec{\Gamma}}{e \equiv e'}{\type}}
\end{mathpar}
\captionof{figure}{The equational judgment, complete, written $\typed{\vec{\Gamma}}{e_1 \equiv e_2}{\type}$.
\textsc{beta-prod} and \textsc{beta-case} are shown for $\syn{L}$ and $\syn{inl}$; the $\syn{R}$/$\syn{inr}$ versions are dual, and \textsc{cong-forall} has an $\exists$ dual.
Not shown: pointwise congruence rules in an unchanged context for $\syn{app}$, $\odot$, $\syn{pure}$, pairing, projections, injections, $\syn{case}$, the connectives, equality, and $\syn{lift}$. 
In \textsc{beta-embed-apply}, $f \col T \to S$ and $x \col T$ are Lean values.
In \textsc{ax}, the third premise is equality of the two interpretations in the model.
}
\label{fig:eq-full}
\end{figrules}
\bigskip
 
\end{document}